\documentclass[preprint, 10pt, numbers]{sigplanconf}

\usepackage{tikz}         
\usepackage{caption}
\DeclareCaptionType{copyrightbox}
\usepackage{pgfplots}
\usepackage{amssymb}      
\usepackage{pifont}       
\usepackage{url}          
\usepackage{fancyvrb}     
\usepackage{listings}
\usepackage{paralist}
\usepackage{color}                

\newcommand{\co}[1]{\lstinline[breaklines=true,breakatwhitespace=true]{#1}}
\newcommand{\mkcol}[1]{\textcolor{red}{\textbf{#1}}}

\begin{document}

\title{Verification of the Tree-Based Hierarchical Read-Copy~Update in the Linux Kernel}
\date{}


\authorinfo{Lihao Liang}
           {University of Oxford}

\authorinfo{Paul E. McKenney}
           {IBM Linux Technology Center}

\authorinfo{Daniel Kroening \\ Tom Melham}
           {University of Oxford}



\maketitle

\begin{abstract}
Read-Copy Update (RCU) is a scalable, high-performance Linux-kernel
synchronization mechanism that runs low-over\-head readers concurrently 
with updaters.
Pro\-duc\-tion-quality RCU implementations for multi-core systems are 
decidedly non-trivial. Giving the ubiquity of Linux, a rare
``million-year'' bug can occur several times per day across the installed
base.
Stringent validation of RCU's complex behaviors is thus critically important. 
Exhaustive testing is infeasible due to the exponential
number of possible executions, which suggests use of formal verification.

Previous verification efforts on RCU either focus on simple implementations 
or use modeling languages, the latter requiring
error-prone manual translation that must be repeated frequently
due to regular changes in the Linux kernel's RCU implementation.
In this paper, we first describe the implementation of Tree RCU in the 
Linux kernel. We then discuss how to construct a model directly from Tree RCU's 
source code in C, and use the CBMC model checker to verify its safety and liveness properties.
To our best knowledge, this is the first verification of a significant part 
of RCU's source code, and is an important step towards integration of formal verification into 
the Linux kernel's regression test suite.
\end{abstract}

\category{}{D.2.4}{Software/Program Verification}[Model checking]
\category{}{D.1.3}{Concurrent Programming}[Parallel programming]

\keywords
Software Verification, Parallel Computing, Read-Copy Update, Linux Kernel


\section{Introduction}
The Linux operating system kernel~\cite{LinuxKernel} is widely used 
in a variety of computing platforms, including servers, safety-critical 
embedded systems, household appliances, and mobile devices such as 
smartphones. Over the past 25 years, many technologies 
have been added to the Linux kernel, one example being Read-Copy 
Update (RCU)~\cite{McKenneyRCU98}.

RCU is a synchronization mechanism that can be used to replace reader-writer 
locks in read-mostly scenarios. It allows low-overhead readers
to run concurrently with updaters. Production-quality RCU implementations 
for multi-core systems must provide excellent scalability, 
high throughput, low latency, modest memory footprint, excellent energy 
efficiency, and reliable response to CPU hotplug operations.
The implementation must therefore avoid cache misses, lock contention, 
frequent updates to shared variables, and excessive 
use of atomic read-modify-write and memory-barrier instructions.
Finally, the implementation must cope with the extremely diverse
workloads and platforms of Linux~\cite{McKenneyOSR08}.

RCU is now widely used in the Linux-kernel networking, device-driver, and
file-storage
subsystems~\cite{McKenneyOSR08,McKenneyRCUsageReport13}.
To date, there are at least 75 million Linux servers \cite{LinuxServerNum13} 
and 1.4 billion Android devices~\cite{AndroidNum15},
which means that a ``million-year'' bug
can occur several times per day across the installed base.
Stringent validation of RCU's complex implementation is thus critically
important.

Most validation efforts for concurrent software rely on testing, but
unfortunately there is no cost-effective test strategy that can cover
all corner cases. 
Worse still, some of errors that testing does detect might be difficult to 
reproduce, diagnose, and repair. The concurrent nature of RCU and the sheer 
size of the search space suggest use of formal verification, particularly
model checking~\cite{BurchInfComput92}.

Formal verification has already been applied to some aspects of RCU design, 
including Tiny RCU~\cite{VerificationChallenges}, userspace RCU~\cite{DesnoyersOSR13}, 
sysidle~\cite{VerificationChallenges}, and interactions between
dyntick-idle and non-maskable interrupts (NMIs)~\cite{ValDyntickNMI}.
%
But these efforts either validate trivial single-CPU RCU implementations
in~C (Tiny RCU), or use special-purpose languages such as
Promela~\cite{HolzmannTSE97SPIN}.  Although special-purpose modeling
languages do have advantages, a major disadvantage in the context of the
Linux kernel is the difficult and error-prone translation from source code.
Other researchers have applied manual proofs in formal logics to simple 
RCU implementations~\cite{YangESOP13RCU,DreyerPLDI15RCU}. These proofs are 
quite admirable, but require even more manual work, in addition
to the translation effort.

Worse yet, Linux kernel releases are only about 60 days apart, and RCU
changes with each release.  Thus, any manual work must be replicated about
six times a year so that mechanical and manual models or proofs remain
synchronized with the Linux-kernel RCU implementation.
Therefore, if formal verification is to be part of Linux-kernel RCU's 
regression suite, the verification methods must be scalable and automated. 
%
To this end, this paper describes how to build a model directly from the
Linux kernel source code, and use the C~Bounded Model Checker
(CBMC)~\cite{KroeningTACAS04CBMC} to verify RCU's safety and liveness
properties.
To the best of our knowledge, this is the first automatic
verification of a significant part of the Linux-kernel RCU source code.
\section{Background}

\subsection{What is RCU?}

Read-copy update (RCU) is a synchronization mechanism
that is often used to replace reader-writer locking.
RCU readers run concurrently with updaters,
and so RCU avoids read-side contention by maintaining multiple versions of
objects and ensuring 
they are not freed until all pre-existing readers complete, that is,
until after a \emph{grace period} elapses.
The basic idea is to split updates into removal
and reclamation phases that are separated by a grace period~\cite{McKenneyRCU98}.
The removal phase removes reader-accessible references 
to objects, perhaps by replacing them with new versions.

Modern CPUs guarantee that writes to single aligned
pointers are atomic, so that readers
see either the old or new version of the data structure.
These atomic-write semantics enable atomic
insertions, deletions, and replacements in a linked structure.
This in turn enables readers to dispense with expensive
atomic operations, memory barriers, and cache misses.
In fact, in the most aggressive configurations of Linux-kernel RCU, readers
can use exactly the same sequence of instructions that would be used in
a single-threaded implementation, providing
RCU readers with excellent performance and scalability.

\begin{figure}[tbp]
\centering
\includegraphics[scale=0.35]{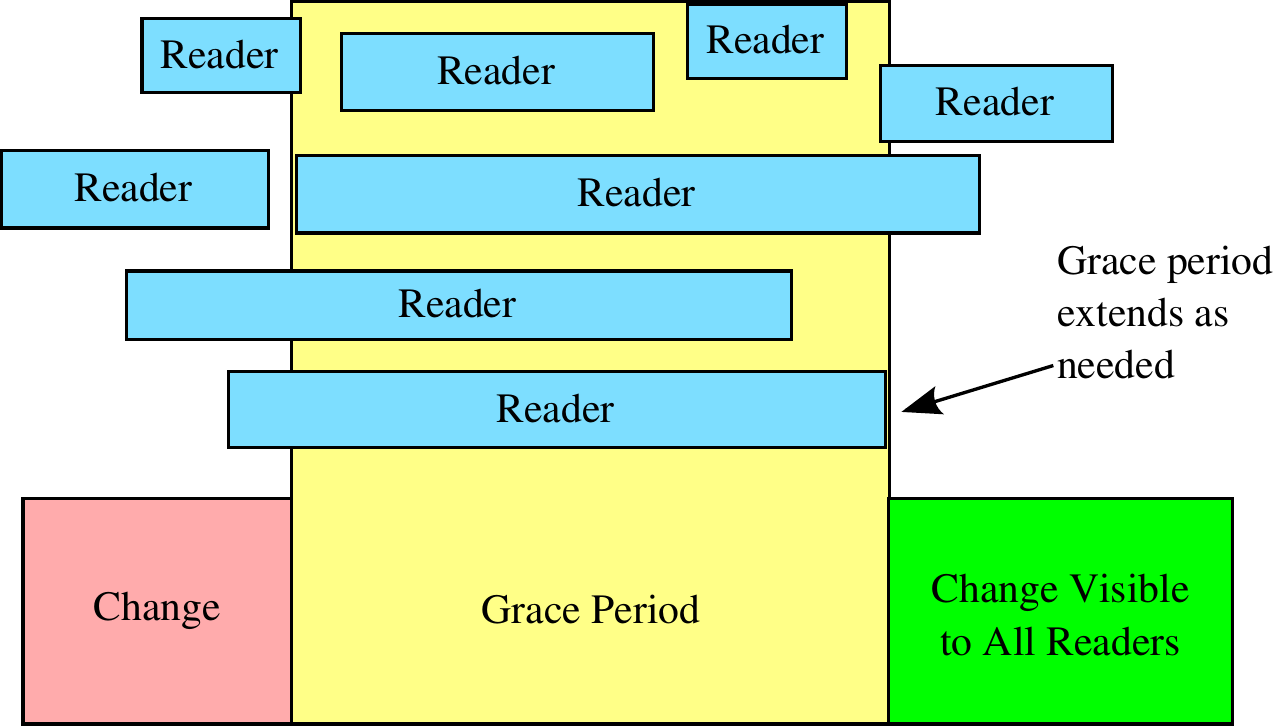}
\caption{RCU Concepts}
\label{fig:rcu_concepts}
\end{figure}

As illustrated in Figure~\ref{fig:rcu_concepts}, grace periods are only
needed for those readers whose runtime overlaps the removal phase.
Those that start after the removal 
phase cannot hold references to the removed objects and thus
cannot be disrupted by objects being freed during the reclamation phase.

\subsection{Core RCU API Usage} \label{sec:api_usage}
The core RCU API is quite small and consists of only five primitives:
\co{rcu_read_lock()}, \co{rcu_read_unlock()}, \co{synchronize_rcu()},
\co{rcu_assign_pointer()}, and \co{rcu_dereference()}~\cite{McKenneyOSR08}.

An RCU read-side critical section begins with \co{rcu_read_lock()}
and ends with a corresponding \co{rcu_read_unlock()}.
When nested, they are flattened into one large critical section.
Within a critical section, it is illegal to block, but
preemption is legal in a preemptible kernel.
RCU-protected data accessed by a 
read-side critical section will not be reclaimed until after that
critical section completes.

The function \co{synchronize_rcu()} marks the end of the updater code and 
the beginning of the reclaimer code. It blocks until all pre-existing RCU
read-side critical sections have completed. Note that \co{synchronize_rcu()} 
does not necessarily wait for critical sections that begin after
it does.


\begin{figure}[tbp]
\centering
\footnotesize
\begin{verbatim}
               int x = 0;
               int y = 0;
               int r1, r2;
               
               void rcu_reader(void) {
                 rcu_read_lock();
                 r1 = x; 
                 r2 = y; 
                 rcu_read_unlock();
               }
               
               void rcu_updater(void) {
                 x = 1; 
                 synchronize_rcu();
                 y = 1; 
               }

               ...

               // after both rcu_reader() 
               // and rcu_updater() return
               assert(r2 == 0 || r1 == 1);
\end{verbatim}
\caption{Verifying RCU Grace Periods}
\label{fig:verify_rcu_gp}
\end{figure}

Consider the example given in
Figure~\ref{fig:verify_rcu_gp}.
If the RCU read-side critical section in function \co{rcu_reader()} begins
before \co{synchronize_rcu()} in \co{rcu_updater()} is called, then it must 
finish before \co{synchronize_rcu()} returns, so that the value of
\co{r2} must be 0. If it ends after \co{synchronize_rcu()} returns, then 
the value of \co{r1} must be 1.

Finally, to assign a new value to an RCU-protected pointer, RCU updaters use 
\co{rcu_assign_pointer()}, which returns the new value. RCU readers 
can use \co{rcu_dereference()} to fetch an RCU-protected pointer, 
which can then be safely dereferenced. 
The returned value is only valid within the enclosing RCU read-side critical section.
The \co{rcu_assign_pointer()} and \co{rcu_dereference()} functions work
together to ensure that if a given reader dereferences an RCU-protected pointer to 
a just-inserted object, the dereference operation will return valid data rather
than pre-initialization garbage.

\section{Implementation of Tree RCU}\label{sec:tree_rcu}

The primary advantage of RCU is that it is able to wait for an arbitrarily
large number of readers to finish without keeping track every single one of
them.  The number of readers can be large (up to the number of CPUs in
non-pre\-emptible implementations and up to the number of tasks in
preemptible implementations).
Although RCU's read-side primitives enjoy excellent performance and scalability, 
update-side primitives must defer the reclamation phase till all pre-existing 
readers have completed, either by blocking or by registering a callback that is 
invoked after a grace period. The performance and scalability of RCU relies on 
efficient mechanisms to detect when a grace period has completed. 
For example, a simplistic RCU implementation might require each CPU to
acquire a global lock during each grace period, but this would severely
limit performance and scalability.
Such an implementation would be quite unlikely to scale beyond
a few hundred CPUs.
This is woefully insufficient because Linux runs on systems with thousands
of CPUs.
This has motivated the creation of Tree RCU.

\subsection{Overview}
%
%
%
We focus on the ``vanilla'' RCU API in a non-preemptible build of
the Linux kernel, specifically on the \co{rcu_read_lock()},
\co{rcu_read_unlock()}, and \co{synchronize_rcu()} primitives.
The key idea is that RCU read-side primitives are confined to kernel code and,
in non-pre\-emptible implementations, do not block.
Thus, when a CPU is blocking, in the idle loop, or running in user mode,
all RCU read-side critical sections that were previously running on that CPU
must have finished. Each of these states is therefore called a \emph{quiescent state}. 
After each CPU has passed through a quiescent state, the corresponding RCU grace period ends.
The key challenge is to determine when all necessary quiescent
states have completed for a given grace period---and to do so with
excellent performance and scalability.

For example, if RCU used a single data structure to record each CPU's
quiescent states, the result would be extreme lock contention on large systems,
in turn resulting in poor performance and abysmal scalability.
Tree RCU therefore instead uses a tree hierarchy of data structures, 
each leaf of which records quiescent states of a single CPU and propagates the information 
up to the root. When the root is reached, a grace period has ended. Then the grace-period
information is propagated down from the root to the leaves of the tree.
Shortly after the leaf data 
structure of a CPU receives this information, \co{synchronize_rcu()} will return.

In the remainder of this section, we discuss the implementation of the non-pre\-empt\-ible 
Tree RCU in the Linux kernel version 4.3.6. We first briefly discuss the implementation of 
read/write-side primitives.
We then explain Tree RCU's hierarchical data structure which records quiescent states while 
maintaining bounded lock contention. Finally, we discuss how RCU uses this data structure 
to detect quiescent states and 
grace periods without individually tracking readers.

\subsection{Read/Write-Side Primitives} \label{sec:api_impl}
In a non-preemptible kernel, any region of kernel code that does not voluntarily
block is implicitly an RCU read-side critical section.
Therefore, the implementations of \co{rcu_read_lock()} and
\co{rcu_read_unlock()} need do nothing at all, and in fact
in production kernel builds that do not have debugging enabled,
these two primitives have absolutely no effect on code generation.


%
In the common case where there are multiple CPUs running,
%
the update-side primitive \co{synchronize_rcu()} calls
\co{wait_rcu_gp()}, which is an internal function that uses
a callback mechanism to invoke \co{wakeme_after_rcu()}
at the end of some later grace period.
As its name suggests, \co{wakeme_after_rcu()} function wakes up
\co{wait_rcu_gp()}, which returns, in turn allowing
\co{synchronize_rcu()} to return control to its caller.

\subsection{Data Structures of Tree RCU} \label{sec:data_structure}

\begin{figure}[tbp]
\centering
\includegraphics[scale=0.2]{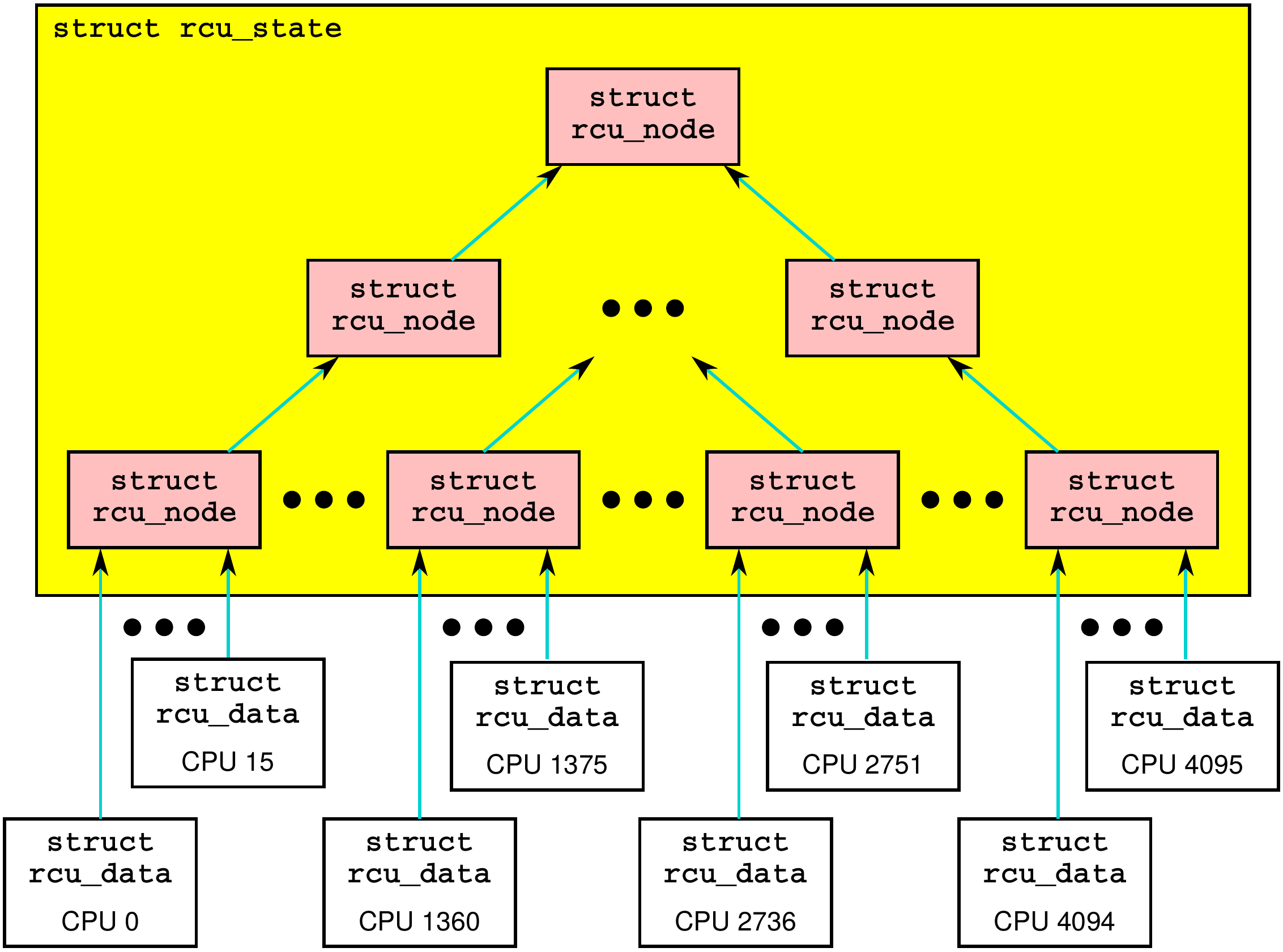}
\caption{Tree RCU Hierarchy}
\label{fig:tree_rcu_hierarchy}
\end{figure}

RCU's global state is recorded in the \co{rcu_state} structure, which consists of 
a tree of \co{rcu_node} structures with a child count of up to 64
(32 in a 32-bit system). Every leaf node can have at most 64 
\co{rcu_data} structures (again 32 on a 32-bit system), each representing
a single CPU, as illustrated in
Figure~\ref{fig:tree_rcu_hierarchy}.
Each \co{rcu_data} structure records its CPU's quiescent states, and
the \co{rcu_node} tree propagates these states up to the root, and then
propagates grace-period information back down to the leaves.
Quiescent-state information does not propagate upwards from a given node
until a quiescent state has been reported by each CPU covered by the subtree
headed by that node.
This propagation scheme dramatically reduces the lock contention experienced
by the upper levels of the tree.
%
For example, consider a default \co{rcu_node} tree for a 4,096-CPU system,
which will have have 256 leaf nodes, four internal nodes, and one root node.
During a given grace period, each CPU will report its quiescent states
to its leaf node, but there will only be 16 CPUs contending for each of
those 256 leaf nodes.
Only 256 of the CPUs will report quiescent states to the internal nodes,
with only 64 CPUs contending for each of the four internal nodes.
Only four CPUs will report quiescent states to the root node, resulting
in extremely low contention on the root node's lock, so that contention
on any given \co{rcu_node} structure is sharply bounded even in very
large configurations.
The current RCU implementation in the Linux kernel supports up to a
four-level tree, and thus in total $64^4 = 16,777,216$ CPUs in a 64
bit machine.\footnote{
	Four-level trees are only used in stress testing,
	but three-level trees are used in production by 4096-CPU systems.}

\subsubsection{\co{rcu_state} Structure}

\begin{figure}[tbp]
\centering
\includegraphics[scale=0.9]{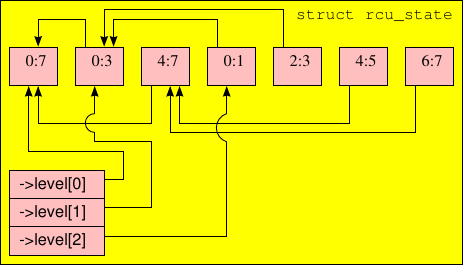}
\caption{Array Representation for a Tree of \co{rcu_node} Structures}
\label{fig:rcu_node_array}
\end{figure}

Each flavor of RCU has its own global \co{rcu_state} structure. 
The \co{rcu_state} structure includes
an array of \co{rcu_node} structures organized as a tree
\co{struct rcu_node node[NUM_RCU_NODES]}, with
\co{rcu_data} structures connected to the leaves.
Given this organization, a breadth-first traversal is 
simply a linear scan of the array.
Another array \co{struct rcu_node} \co{*level[NUM_RCU_LVLS]} 
is used to point to the left-most node at each level of the tree,
as shown in Figure~\ref{fig:rcu_node_array}.

The \co{rcu_state} structure uses \co{unsigned long} fields \co{->gpnum}
and \co{->completed} to track RCU's grace periods.
The \co{->gpnum} field records the most recently started grace period,
whereas \co{->completed} records the most recently ended grace period.
If the two numbers are equal, then corresponding flavor of RCU is idle.
If \co{gpnum} is one greater than \co{completed}, then RCU is in the
middle of a grace period.
All other combinations are invalid.


\subsubsection{\co{rcu_node} Structure}
\label{sec:rcu_node}
The tree of \co{rcu_node} structures records and 
propagates quiescent-state information from the leaves to the root,
and also propagates grace-period information from the root to the leaves. 
The \co{rcu_node} structure has a spinlock \co{->lock} to protect its fields.
The \co{->parent} field references the parent \co{rcu_node} structure,
and is \co{NULL} for the root.
The \co{->level} field indicates the level in the tree, counting from zero
at the root.
The \co{->grpmask} field identifies this node's bit in the
\co{->qsmask} field of its parent.
The \co{->grplo} and \co{->grphi} fields indicates the lowest and highest 
numbered CPU that are covered by this \co{rcu_node} structure, respectively.

The \co{->qsmask} field indicates which of this node's children
still need to report quiescent states for the current grace period.
As with \co{rcu_state}, the \co{rcu_node} structure has \co{->gpnum} 
and \co{->completed} fields that have values identical to those of the
enclosing \co{rcu_state} structure, except at the beginnings and ends
of grace periods when the new values are propagated down the tree.
Each of these fields can be smaller than 
its \co{rcu_state} counterpart by at most one.


\subsubsection{\co{rcu_data} structure} \label{sec:rcu_data}
The \co{rcu_data} structure detects quiescent states and handles RCU
callbacks for the corresponding CPU.
The structure is accessed primarily from the corresponding CPU,
thus avoiding synchronization overhead.
As with the \co{rcu_state} structure, different flavors of RCU maintain 
their own per-CPU \co{rcu_data} structures. 
%
The \co{->cpu} field identifies the corresponding CPU, the \co{->rsp}
field references the corresponding \co{rcu_state} structure, and the
\co{->mynode} field references the corresponding leaf \co{rcu_node}
structure.
The \co{->grpmask} field identifies this \co{rcu_data} structure's bit
in the \co{->qsmask} field of its leaf \co{rcu_node} structure.

The \co{rcu_data} structure's \co{->qs_pending} field indicates that RCU
needs a quiescent state from the corresponding CPU, and the
\co{->passed_quiesce} indicates that the CPU has already passed through
a quiescent state.
The \co{rcu_data} also has \co{->gpnum} and \co{->completed} fields,
which can lag arbitrarily behind their counterparts in
the \co{rcu_state} and \co{rcu_node} structures on idle CPUs.
However, on the non-idle CPUs that are the focus of this paper,
these two fields can lag at most one grace period behind their leaf \co{rcu_node} 
counterparts.

The \co{rcu_state} structure's \co{->gpnum} and \co{->completed} fields
represent the most current values, and are tracked closely by those of
the \co{rcu_node} structure, which allows the \co{->gpnum} and
\co{->completed} fields in the \co{rcu_data} structures to be
are compared against their counterparts in the corresponding leaf \co{rcu_node}
to detect a new grace period. 
This scheme allows CPUs to detect beginnings and ends of grace periods without
incurring lock- or memory-contention penalties.
The \co{rcu_data} structure manages RCU callbacks using a 
four-segment list~\cite{LaiJiangshan2008NewClassicAlgorithm}.


\begin{figure}[tbp]
\centering
\includegraphics[scale=0.25]{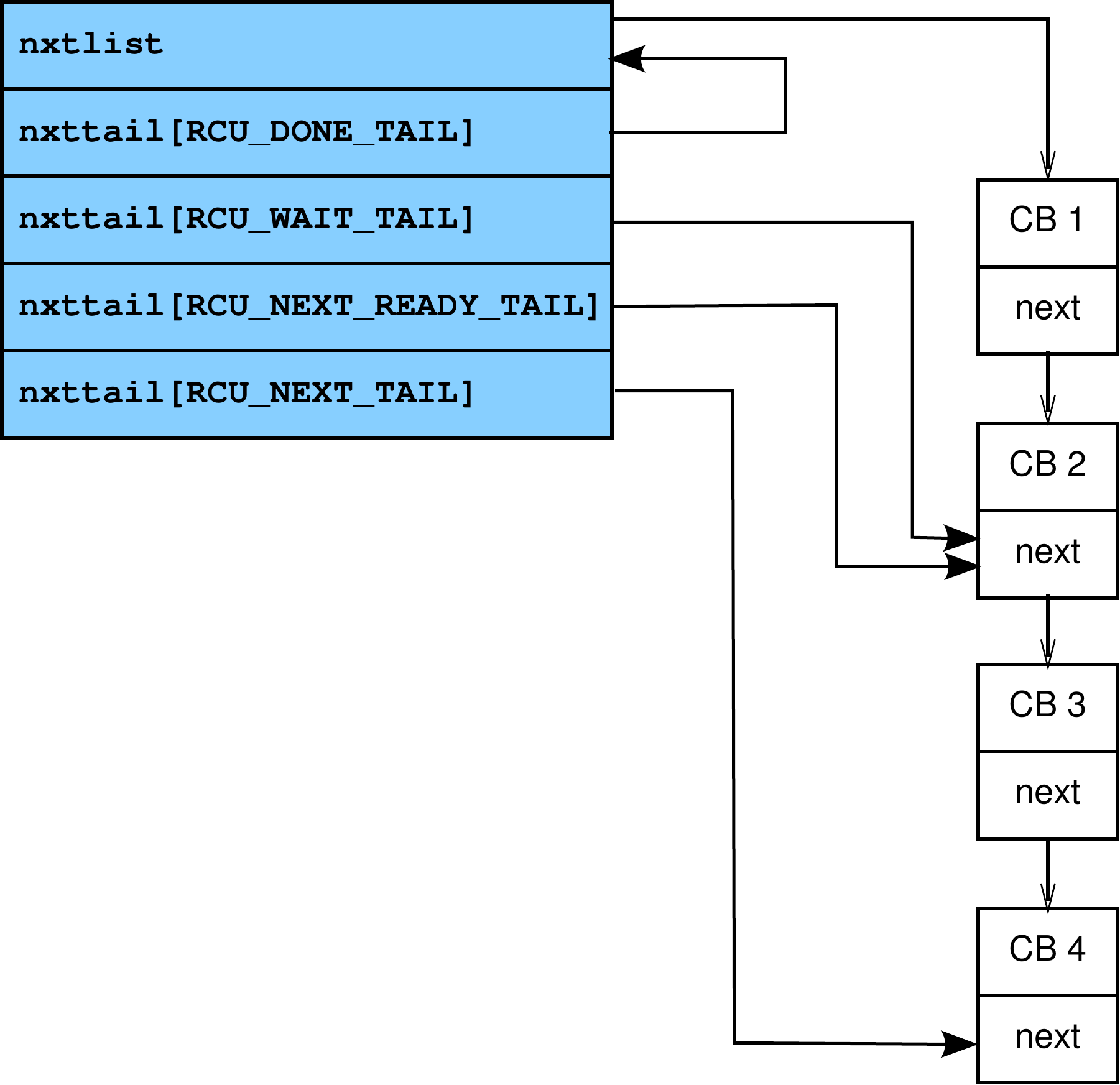}
\caption{Callback Queuing in \co{rcu_data}}
\label{fig:rcu_data_callbacks}
\end{figure}

\subsubsection{RCU Callbacks}
The \co{rcu_data} structure manages RCU callbacks using a \co{->nxtlist}
pointer tracking the head of the list and an array of \co{->nxttail[]}
tail pointers that form a four-segment list of
callbacks~\cite{LaiJiangshan2008NewClassicAlgorithm}, with
each element of the \co{->nxttail[]} array referencing the tail of the
corresponding segment, as shown in Figure~\ref{fig:rcu_data_callbacks}.
The segment ending with \co{->nxttail[RCU_DONE_TAIL]} (the ``\co{RCU_DONE_TAIL}
segment'') contains callbacks
handled by a prior grace period that are therefore ready to be invoked.
The \co{RCU_WAIT_TAIL} and \co{RCU_NEXT_READY_TAIL} segments 
contain callbacks waiting for the
current and the next grace period, respectively.
Finally, the \co{RCU_NEXT_TAIL} segment contains
callbacks that are not yet associated with any grace period.
The \co{->qlen} field counts the total number of callbacks, and
the \co{->blimit} field specifies the maximum number of RCU callbacks
that may be invoked at a given time, thus limiting response-time
degradation due to long lists of callbacks.\footnote{
	Workloads requiring aggressive real-time guarantees should use
	callback offloading, which is outside of the scope of this paper.}

Back in Figure~\ref{fig:rcu_data_callbacks}, the
\co{->nxttail[RCU_DONE_TAIL]} array element references \co{->nxtlist}, 
which means none of the callbacks are ready to invoke.
The \co{->nxttail[RCU_WAIT_TAIL]} element references callback 2's \co{->next}
pointer, meaning that callbacks CB~1 and CB~2 are waiting for the current
grace period.
The \co{->nxttail[RCU_NEXT_READY_TAIL]} element references that same \co{->next}
pointer, meaning that no callbacks are waiting for the next grace period. 
Finally, the callbacks between the \co{->nxttail[RCU_NEXT_READY_TAIL]} and
\co{->nxttail[RCU_NEXT_TAIL]} elements (CB~3 and CB~4)
are not yet assigned to a specific grace period.
The \co{->nxttail[RCU_NEXT_TAIL]} element always references either
the last callback or, when the entire list is empty, \co{->nxtlist}.

Cache locality is promoted by invoking callbacks on the CPU that registered
them.
For example, RCU's update-side primitive 
\co{synchronize_rcu()} appends callback \co{wakeme_after_rcu()} to the end
of the \co{->nxttail[RCU_NEXT_TAIL]} list in the current CPU 
(Section \ref{sec:api_impl}). 
They are advanced one segment towards the head of the list (via \co{rcu_advance_cbs()}) 
when the CPU detects the current grace period has ended, which is indicated 
by the \co{->completed} field of the CPU's \co{rcu_data} structure being one
smaller than its counterpart in the corresponding leaf \co{rcu_node} structure.
The CPU also periodically merges the \co{RCU_NEXT_TAIL} segment into the
\co{RCU_NEXT_READY_TAIL} segment by calling \co{rcu_accelerate_cbs()}.
In a few special cases, the CPU merges the \co{RCU_NEXT_TAIL} segment
into the \co{RCU_WAIT_TAIL} segment, bypassing the \co{RCU_NEXT_READY_TAIL}
segment.
This optimization applies when the CPU is starting a new grace period.
It does \emph{not} apply when a CPU notices a new grace period
because that grace period might well have started before
the callbacks were added to the \co{RCU_NEXT_TAIL} segment.
%
This is a deliberate design choice: It is more important for the CPUs
to operate independently (thus avoiding contention and synchronization
overhead) than it is to decrease grace-period latencies.
In those rare occasions where low grace-period latency is important,
the \co{synchronize_rcu_expedited()} should be used.
This function has the same semantics as does \co{synchronize_rcu()},
but trades off efficiency optimizations in favor of reduced latency.

Each RCU callbacks is an \co{rcu_head} structure which has a
\co{->next} field that points to the next callback on the list and
a \co{->func} field that references the function to be invoked at the
end of an upcoming grace period.

\subsection{Quiescent State Detection} \label{sec:quiescent_state}
RCU has to wait until all pre-existing read-side critical sections have
finished before it can safely allow a grace period to end.
The performance and scalability of RCU rely on its ability to efficiently
detect quiescent states and determine whether the set of quiescent states
detected thus far allows the grace period to end.
If each CPU (or, in the case of preemptible RCU, each task)
has passed through a quiescent state, a grace period has elapsed. 

The non-preemptible RCU-sched flavor's quiescent states
apply to CPUs, and are user-space execution, context switch, idle, and 
offline state.
%
%
Therefore, RCU-sched 
only needs to track tasks and interrupt handlers that are actually running because
blocked and preempted tasks are always in quiescent states. Thus, RCU-sched 
needs only track CPU states.

\subsubsection{Scheduling-Clock Interrupt} \label{sec:timer_interrupt}
The \co{rcu_check_callbacks()} is invoked from the sched\-ul\-ing-clock interrupt
handler, which allows RCU to periodically check whether a given busy CPU
is in the user-mode or idle-loop quiescent states.
If the CPU is in one of these quiescent states, \co{rcu_check_callbacks()}
invokes \co{rcu_sched_qs()}, 
which sets the per-CPU \co{rcu_sched_data.passed_quiesce} 
fields to 1. 

%
The \co{rcu_check_callbacks()} function invokes \co{rcu_pending()}
to determine whether a recent event or current condition means that
RCU requires attention from this CPU.
If so, \co{rcu_check_callbacks()} invokes \co{raise_softirq()}, 
which will cause \co{rcu_process_callbacks()} to be invoked once the CPU 
reaches a state where it is safe to do so (roughly speaking, once the CPU 
has interrupts, preemption, and bottom halves enabled). This function is 
discussed in detail in Section \ref{sec:rcu_softirq}.



\subsubsection{Context-Switch Handling} \label{sec:context_switch}
The context-switch quiescent state is recorded by invoking
\co{rcu_note_context_switch()} from \co{__schedule()} (and, for the
benefit of virtualization, also from \co{rcu_virt_note_context_switch()}).
%
The \co{rcu_note_context_switch()} function invokes \co{rcu_sched_qs()}
to inform RCU of the context switch, which is a quiescent state of the CPU.

\subsection{Grace Period Detection} \label{sec:grace_period}
\begin{figure}[tb]
\centering
\includegraphics[scale=0.25]{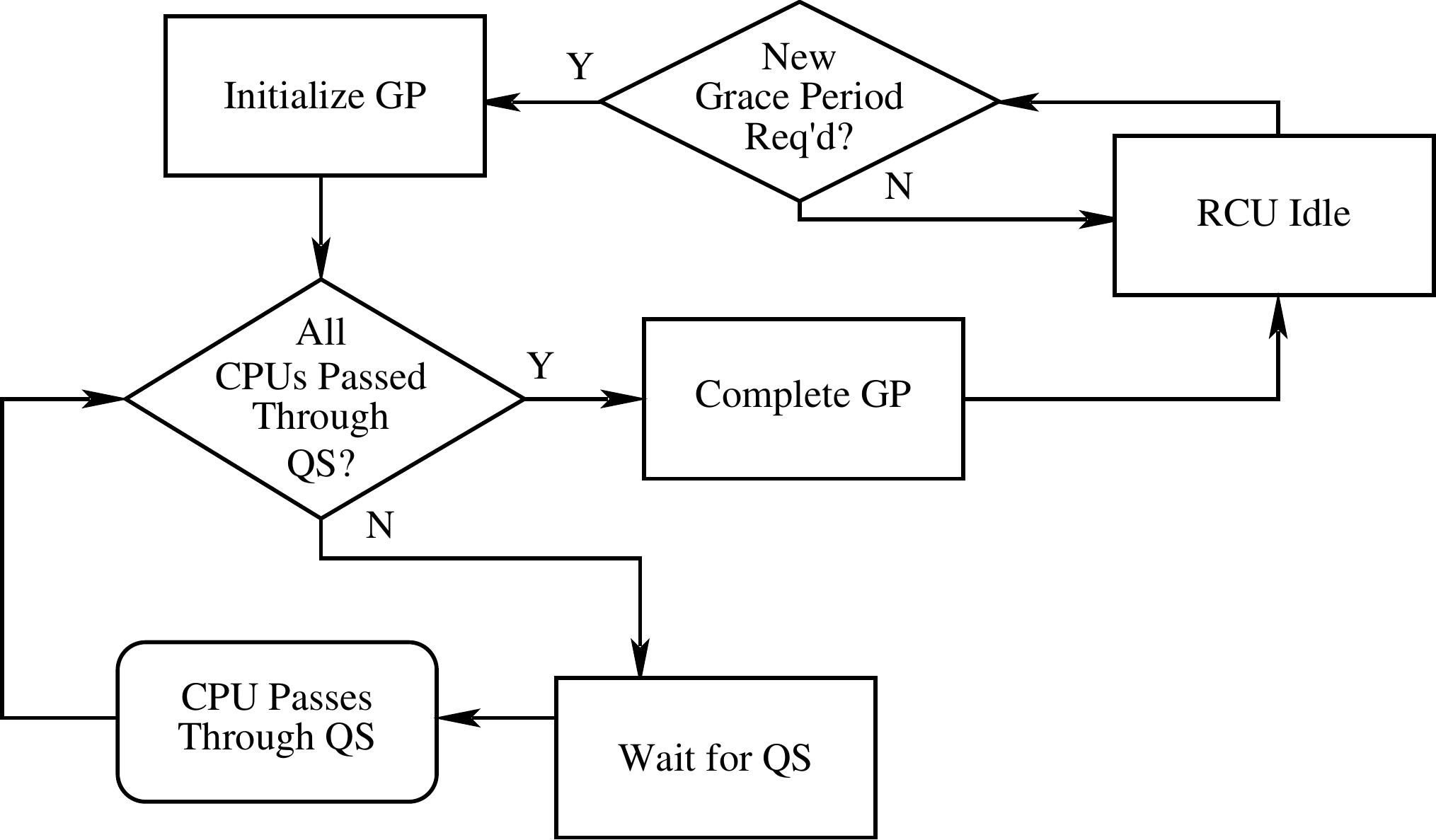}
\caption{Grace-Period Detection State Diagram}
\label{fig:grace_period_state_diagram}
\end{figure}

Once each CPU has passed through a quiescent state, a grace period for RCU
has completed. 
As discussed in Section \ref{sec:data_structure}, Tree-RCU uses a hierarchy 
of \co{rcu_node} structures to manage quiescent state and grace period
information.
Quiescent-state information is passed up the 
tree from the leaf per-CPU \co{rcu_data} structures.
Grace-period information is passed down from the root.
%
We focus on grace-period detection for busy CPUs, as illustrated
in Figure~\ref{fig:grace_period_state_diagram}.

\subsubsection{Softirq Handler for RCU} \label{sec:rcu_softirq}
RCU's busy-CPU grace period detection relies on the
\co{RCU_SOFTIRQ} handler function \co{rcu_process_callbacks()},
which is scheduled from the scheduling-clock interrupt.
This function first calls
\co{rcu_check_quiescent_state()} to report recent quiescent states
on the current CPU.
Then \co{rcu_process_callbacks()} starts a new grace period if needed,
and finally calls \co{invoke_rcu_callbacks()} to invoke any callbacks
whose grace period has already elapsed.

Function \co{rcu_check_quiescent_state()} first invokes \co{note_gp_changes()} 
to update the CPU-local \co{rcu_data} structure to record the end of 
previous grace periods and the beginning of new grace periods.
Any new values for these fields are copied from the leaf \co{rcu_node}
structure to the \co{rcu_data} structure.
If an old grace period has ended, \co{rcu_advance_cbs()} is invoked to
advance all callbacks, otherwise, \co{rcu_accelerate_cbs()} is invoked
to assign a grace period to any recently arrived callbacks.
If a new grace period has started, \co{->passed_quiesce} is set to zero,
and if in addition RCU is waiting for a quiescent state from this CPU,
\co{->qs_pending} is set to one, so that a new quiescent state will
be detected for the new grace period.
%

Next,
\co{rcu_check_quiescent_state()} checks whether \co{->qs_pending} indicates
that RCU needs a quiescent state from this CPU.
If so, it checks whether \co{->passed_quiesce} indicates that this
CPU has in fact passed through a quiescent state.
If so, it invokes \co{rcu_report_qs_rdp()} to report that quiescent
state up the 
combining tree.

The \co{rcu_report_qs_rdp()} function first verifies that the CPU has
in fact detected a legitimate quiescent state for the current grace period,
and under the protection of the leaf \co{rcu_node} structure's \co{->lock}.
If not, it resets quiescent-state detection and returns, thus ignoring
any redundant quiescent states belonging to some earlier grace period.
Otherwise, if the \co{->qsmask} field indicates that RCU needs to report a 
quiescent state from this CPU, \co{rcu_accelerate_cbs()} is invoked to assign 
a grace-period number to any new callbacks, and then \co{rcu_report_qs_rnp()} 
is invoked to report the quiescent state to the \co{rcu_node} combining tree.

%
%
%
%

The \co{rcu_report_qs_rnp()} function traverses up the \co{rcu_node} tree,
at each level holding the \co{rcu_node} structure's \co{->lock}.
At any level, if the child structure's \co{->qsmask} bit is already clear,
or if the \co{->gpnum} changes, traversal stops.
Otherwise, the child structure's bit is cleared from \co{->qsmask},
after which, if \co{->qsmask} is non-zero, 
traversal stops. Otherwise, traversal proceeds on to the parent \co{rcu_node} structure.
Once the root is reached, traversal stops and \co{rcu_report_qs_rsp()} is
invoked to awaken the grace-period kthread (kernel thread).
The grace-period kthread will then clean up after the now-ended grace
period, and, if needed, start a new one.

\subsubsection{Grace-Period Kernel Thread} \label{sec:rcu_gp_kthread}
The RCU grace-period kthread invokes \co{rcu_gp_kthread()}, which
contains an infinite loop that initializes, waits for, and cleans up after
each grace period. 

When no grace period is required, the grace-period kthread
sets its \co{rcu_state} structure's \co{->flags} field to
\co{RCU_GP_WAIT_GPS}, and then
waits within an inner infinite loop for that structure's
\co{->gp_state} field to be set.
Once set, \co{rcu_gp_kthread()} invokes \co{rcu_gp_init()} to initialize
a new grace period, which
rechecks the \co{->gp_state} field under
the root \co{rcu_node} structure's \co{->lock}.
If the field is no longer set, \co{rcu_gp_init()} returns zero.
Otherwise, it
increments \co{rsp->gpnum} by 1 to record a new grace period number.
Finally, it performs a breadth-first traversal of the \co{rcu_node}
structures in the combining tree.
For each \co{rcu_node} structure \co{rnp},
%
we set the \co{rnp->qsmask} to indicate which children
must report quiescent states for the new grace period (Section 
\ref{sec:rcu_node}), and set \co{rnp->gpnum} and \co{rnp->completed}
to their \co{rcu_state} counterparts. 
If the \co{rcu_node} structure \co{rnp} is the parent of the current CPU's \co{rcu_data}, 
we invoke \co{__note_gp_changes()} to set up the CPU-local \co{rcu_data} state. 
Other CPUs will invoke \co{__note_gp_changes()} after their next
scheduling-clock interrupt. 
 


To clean up after a grace period, \co{rcu_gp_kthread()} 
calls \co{rcu_gp_cleanup()} after setting the \co{rcu_state} field \co{rsp->gp_state} 
to \co{RCU_GP_CLEANUP}. After the function returns, \co{rsp->gp_state} is set to 
\co{RCU_GP_CLEANED} to record the end of the old grace period.
Function \co{rcu_gp_cleanup()} performs a breadth-first traversal of
\co{rcu_node} combining-tree.
It first sets each \co{rcu_node} structure's \co{->completed} field
to the \co{rcu_state} structure's \co{->gpnum} field.
It then updates the current CPU's CPU-local \co{rcu_data} structure by
calling \co{__note_gp_changes()}. 
For other CPUs, the update will take place when they handle the scheduling-clock
interrupts, in a fashion similar to \co{rcu_gp_init()}. 
After the traversal, it marks the completion of the grace period by setting the
\co{rcu_state} structure's \co{->completed}
field to that structure's \co{->gpnum} field, and invokes
\co{rcu_advance_cbs()} to advance callbacks. 
Finally, if another grace period is needed,
we set \co{rsp->gp_flags} to \co{RCU_GP_FLAG_INIT}. 
Then in the next iteration of the outer loop, the grace-period kthread
will initialize a new grace period as discussed above.




\section{Verification Scenario}

We use the example in Figure~\ref{fig:verify_rcu_gp} to demonstrate how the
different components of Tree RCU work together to guarantee that all
pre-existing read-side critical sections finish before RCU allows a grace
period to end.  This example will drive the verification, which will check
for violations of the assertion at this end of the code.

We focus on the implementation of the non-preemptible RCU-sched flavor.  We
further assume there are only two CPUs, and that CPU~0 executes function
\co{rcu_reader()} and CPU~1 executes \co{rcu_updater()}.  When the system
boots, the Linux kernel calls \co{rcu_init()} to initialize RCU, which
includes constructing the combining tree of \co{rcu_node} and \co{rcu_data}
structures via \co{rcu_init_geometry()} and initializing the fields of the
nodes in the tree for each RCU flavor via \co{rcu_init_one()}.  In our
example it will be a one-level tree that has one \co{rcu_node} structure as
root and two children that are \co{rcu_data} structures for each CPU. 
Function \co{rcu_spawn_gp_kthread()} is also called to initialize and spawn
the RCU grace-period kthread for each RCU flavor.

Referring again to Figure~\ref{fig:verify_rcu_gp},
suppose that \co{rcu_reader()} begins
execution on CPU~0 while \co{rcu_updater()} concurrently sets \co{x} to 1
and then invokes \co{synchronize_rcu()} on CPU~1.
As discussed in Section \ref{sec:api_impl}, \co{synchronize_rcu()}
invokes \co{wait_rcu_gp()}, which in turn registers an RCU callback
that will invoke \co{wakeme_after_rcu()} some time after \co{rcu_reader()}
exits its critical section.

However, this critical-section exit has no immediate effect.
Instead, a later context switch will invoke
\co{rcu_note_context_switch()}, which in turn invokes
\co{rcu_sched_qs()}, recording the quiescent state in the
CPU's \co{rcu_sched_data} structure's \co{->passed_quiesce} field.
Later, a scheduling-clock interrupt will invoke
\co{rcu_check_callbacks()}, which calls \co{rcu_pending()} and 
notes that the \co{->passed_quiesce} field is set.
This will cause \co{rcu_pending()} to return \co{true}, which
in turn causes \co{rcu_check_callbacks()} to invoke
\co{rcu_process_callbacks()}.
In its turn, \co{rcu_process_callbacks()} will invoke
\co{raise_softirq(RCU_SOFTIRQ)}, which,
once the CPU has interrupts, preemption, and
bottom halves enabled, 
calls \co{rcu_process_callbacks()}.

As discussed in Section \ref{sec:rcu_softirq}, RCU's softirq handler function \co{rcu_process_callbacks()} 
first calls \co{rcu_check_quiescent_state()} to report any recent quiescent states on the 
current CPU (CPU~0). Then it checks whether the CPU~0 has passed a quiescent state. Since 
a quiescent state has been recorded for CPU~0, \co{rcu_report_qs_rnp()} is invoked to traversal
up the combining tree. It clears the first bit of the root \co{rcu_node} structure's \co{qsmask} 
field (recall that the RCU combining tree has only one level). Since the second bit for CPU~1 has 
not been cleared, the function returns.

Since \co{synchronize_rcu()} blocks in CPU~1, it will result in a context switch. 
This triggers a sequence of events similar to that described above for
CPU~1, which results in the clearing of the
second bit of the root \co{rcu_node} structure's \co{->qs_mask} field, the value of which is now 0, indicating the end of the current grace period.
CPU~1 therefore invokes \co{rcu_report_qs_rsp()} to 
awaken the grace-period kthread, 
which will clean up the ended grace period, and, if needed, 
start a new one (Section \ref{sec:rcu_gp_kthread}).

Lastly, \co{rcu_process_callbacks()} calls \co{invoke_rcu_callbacks()} to invoke any callbacks whose
grace period has already elapsed, for example, \co{wakeme_after_rcu()},
which will allow \co{synchronize_rcu()} to return.


\section{Modeling RCU for CBMC} \label{sec:model_rcu}

The C Bounded Model Checker
(CBMC)\footnote{\url{http://www.cprover.org/cbmc/}} is a program analyzer
that implements bit-precise bounded model checking for C programs~\cite{KroeningTACAS04CBMC}.
CBMC can demonstrate violation of assertions in C programs, or prove their 
safety under a given loop unwinding bound.
It translates an input C program into a formula, which is then passed to a
modern SAT or SMT solver together with a constraint that specifies the set
of error states.  If the solver determines the formula to be satisfiable, an
error trace giving the exact sequence of events is extracted from the
satisfying assignment.
Recently, support has been added for verifying concurrent programs over a
wide range of memory models, including SC, TSO, and PSO~\cite{AlglaveCAV13}.

In the remainder of this section we describe how to construct a model from
the source code of the Tree RCU implementation in the Linux kernel version 4.3.6, 
which can be verified by CBMC.  Model construction entailed stubbing
out calls to other parts of the kernel, removing irrelevant functionality
(such as idle-CPU detection), removing irrelevant data (such as statistics),
and adding preprocessor directives to conditionally inject bugs (described
in Section~\ref{sec:bug_cases}). 
The Linux kernel environment and the majority of these changes to the source code 
are made through macros in separate files that can be reused across different versions 
of the Tree RCU implementation. The biggest change in the source files is to use arrays 
to model per-CPU data, which could potentially be scripted~\cite{RoyValSRCU17}.
The resulting model has 8,626 lines of C code. Around 900 lines of the code models the
Linux kernel environment. This is significantly smaller than the actual Linux kernel
code used by the Tree RCU implementation, making verification of Tree RCU possible.
The model contains assertions and can be also run as a user program, which provides
important validation of the model itself. 
%
%

\subsubsection*{Initialization}

Our model first invokes \co{rcu_init()} which in turn invokes:
(1)~\co{rcu_init_geometry()} to compute the \co{rcu_node} tree geometry;
(2)~\co{rcu_init_one} to initialize the \co{rcu_state}
structure; (3)~\co{rcu_cpu_notify()} to initialize each CPU's
\co{rcu_data} structure.
This boot initialization tunes the data-structure configuration
to match that of the specific hardware at hand.
For example, a large-system tree might resemble
Figure~\ref{fig:tree_rcu_hierarchy}, while
a small configuration has a single \co{rcu_node} ``tree''.
The model then calls \co{rcu_spawn_gp_kthread()} 
to spawn the grace-period kthreads discussed below.

\subsubsection*{Per-CPU Variables and State}

RCU uses per-CPU data to provide cache locality and to reduce 
contention and synchronization overhead.
For example, the per-CPU structure 
\co{rcu_data} records quiescent states 
and handles RCU callbacks (Section \ref{sec:rcu_data}). 
We model this per-CPU data as an array, indexed by CPU ID.

It is also necessary to model per-CPU state, including the currently
running task and whether or not interrupts are enabled.
Identifying the running task requires a (trivial)
model of the Linux-kernel scheduler, which
uses an integer array \co{cpu_lock}, indexed by CPU ID.
Each element of this array models an exclusive lock.
When a task schedules on a given CPU, it acquires the corresponding CPU lock,
and releases it when scheduling away.
We currently do not model preemption, so need model only voluntary context
switches.

A pair of integer arrays \co{local_irq_depth} and \co{irq_lock} is used
to model CPUs enabling and disabling interrupts.
Both arrays are indexed by CPU ID, with the first recording each CPU's
interrupt-disable nesting depth and the second recording whether
or not interrupts are disabled.

\subsubsection*{Update-Side API \co{synchronize_sched()}}
Because our model omits CPU hotplug and callback handling, we cannot use
Tree RCU's normal callback mechanisms to detect the end of a grace period.
We therefore use a global variable \co{wait_rcu_gp_flag}, which is
initialized to~1 in \co{wait_rcu_gp()} before the grace period.
Because \co{wait_rcu_gp()} blocks, it can result in a context switch,
the model invokes \co{rcu_note_context_switch()}, followed by a call 
to \co{rcu_process_callbacks()} to inform RCU of the resulting
quiescent state.
When the resulting quiescent states propagate to
the root of the combining tree, the grace-period kthread is awakened.
This kthread then invokes \co{rcu_gp_cleanup()}, the modeling of which 
is described below. Then \co{rcu_gp_cleanup()} calls \co{rcu_advance_cbs()}, 
which invokes \co{pass_rcu_gp()} to clear the \co{wait_rcu_gp_flag} flag.
The \co{__CPROVER_assume(wait_rcu_gp_flag == 0)}
~in~ \co{wait_rcu_gp()} prevents CBMC from continuing execution until
\co{wait_rcu_gp_flag} is equal to~0, thus modeling the needed grace-period
wait.

\subsubsection*{Scheduling-Clock Interrupt and Context Switch} \label{sec:model_irq}

The \co{rcu_check_callbacks()} function detects idle and usermode execution, 
as well as invokes RCU core processing in response to state changes.
Because we model neither idle nor usermode execution,
the only state changes are context-switches and the beginnings and ends of grace periods.
We therefore dispense with \co{rcu_check_callbacks()} (Section \ref{sec:timer_interrupt}).
Instead, we directly call \co{rcu_note_context_switch()} just after
releasing a CPU, which in turn calls \co{rcu_sched_qs()} to record the
quiescent state.
Finally, we call \co{rcu_process_callbacks()},
which notes grace-period beginnings and ends and reports quiescent states
up RCU's combining tree.

\subsubsection*{Grace-Period Kthread}
As discussed in Section \ref{sec:rcu_gp_kthread}, \co{rcu_gp_kthread()}
invokes \co{rcu_gp_init()}, \co{rcu_gp_fqs()}, and \co{rcu_gp_cleanup()}
to initialize, wait for, and clean up after each grace period, respectively.
To reduce the size of the formula generated by CBMC, instead of spawning a
separate thread, we directly call \co{rcu_gp_init()} from
\co{rcu_spawn_gp_kthread} and \co{rcu_gp_cleanup()} from
\co{rcu_report_qs_rsp()}.
Because we model neither idle nor usermode execution, we need not
call \co{rcu_gp_fqs()}.

\subsubsection*{Kernel Spin Locks}
CBMC's
\co{__CPROVER_atomic_begin()}, \co{__CPROVER_atomic_end()}, and
\co{__CPROVER_assume()} built-in primitives are used to construct atomic test-and-set
for \co{spinlock_t} and \co{raw_spinlock_t} acquisition and
atomic reset for release.
We use GCC atomic builtins for user-space execution:
\co{while (__sync_lock_test_and_set(lock, 1))} acquires a
lock and \co{__sync_lock_release(lock)} releases it.


\subsubsection*{Limitations}
We model only the fundamental components of Tree RCU, excluding, for example,
quiescent-state forcing, grace-period expediting, and callback handling.
In addition, we make the assumption that all CPUs are busy executing RCU related tasks. 
As a result, we do not model the following scenarios: 1.~CPU hotplug and dyntick-idle; 
2.~Thread-migration failure modes in the Linux kernel involving per-CPU variables; 
3.~RCU priority boosting. 
Nonetheless, we model real-world server-class RCU code paths and data layout on
systems with up to 16 CPUs (default configurations) or up to either 32 or 64 CPUs
(non-default configurations on either 32-bit or 64-bit CPUs).
Moreover, we model scheduling-clock interrupts as 
direct function calls, which, as discussed later, results in failures to model one of 
the bug-injection scenarios. Lastly, the test harness we use only passes through a 
single grace period, so cannot detect failures involving multiple grace periods.
\section{Experiments}

In this section we discuss our experiments verifying the Tree RCU 
implementation in the Linux-kernel. We first describe several bug-injection
scenarios used in the experiments. Next, we report results of user-space runs
of the RCU model. 
Finally, we discuss the verification results using CBMC.
We performed our experiments on a 64-bit machine running Linux 3.19.8
with eight Intel Xeon 3.07\,GHz cores and 48\,GB of memory.
We have made the source code of our RCU model and the experimental data
available at \url{https://github.com/lihaol/verify-treercu}.

\subsection{Bug-Injection Scenarios} \label{sec:bug_cases}
Because we model non-preemptible Tree RCU, each CPU runs exactly one RCU task
as a separate thread.
Upon completion, each task increments a global counter \co{thread_cnt},
enabling the parent thread to verify the completion of all RCU tasks
using a statement \co{__CPROVER_assume(thread_cnt == 2)}.
The base case uses the example in Figure~\ref{fig:verify_rcu_gp}, including
its assertion \co{assert(r2 == 0 || r1 == 1)}.
This assertion does not hold when
RCU's fundamental safety guarantee is violated:
read-side critical sections cannot span grace periods.
We also verify a \emph{weak form} of liveness by inserting an \co{assert(0)} 
after the \co{__CPROVER_assume(thread_cnt == 2)} statement.
This assertion cannot hold, and so it will be violated if at least one grace period completes.
Such a ``verification failure" is in fact the expected behavior for a correct RCU implementation. 
On the other hand, if the assertion is not violated, grace periods never complete, which indicates a liveness bug.

To validate our verification, we also run CBMC with the bug-injection scenarios 
described below,\footnote{Source code: \url{https://www.kernel.org/pub/linux/kernel/v4.x/linux-4.3.6.tar.xz}, directory \co{kernel/rcu}.}
which are simplified versions of bugs encountered in actual practice.
Bugs~2--6 are liveness checks and thus use the
aforementioned \co{assert(0)}, and the remaining scenarios are
safety checks which thus use the base-case assertion in
Figure~\ref{fig:verify_rcu_gp}.

\paragraph*{Bug 1}
This bug-injection scenario makes
the RCU update-side primitive \co{synchronize_rcu()} return immediately 
(line 523 in \co{tree_plugin.h}).
With this injected bug, updaters never wait for readers, which should
result in a safety violation, thus preventing
Figure~\ref{fig:verify_rcu_gp}'s assertion from holding.

\paragraph*{Bug 2}
The idea behind this bug-injection scenario is to prevent individual
CPUs from realizing that quiescent states are needed, thus preventing
them from recording quiescent states. As a result, it prevents grace periods
from completing.
Specifically, in function \co{rcu_gp_init()}, for each \co{rcu_node}
structure in the combining tree,
we set the field \co{rnp->qsmask} to 0 instead of \co{rnp->qsmaskinit} (line 1889 in \co{tree.c}). 
Then when \co{rcu_process_callbacks()} is called, \co{rcu_check_quiescent_state()} will invoke
\co{__note_gp_changes()} that sets \co{rdp->qs_pending} to 0, indicating that RCU needs no 
quiescent state from the corresponding CPU. Thus, \co{rcu_check_quiescent_state()} will return 
without calling \co{rcu_report_qs_rdp()}, preventing grace periods from completing.
This liveness violation should fail to trigger a violation of the end-of-execution
\co{assert(0)}.

\paragraph*{Bug 3}
This bug-injection scenario is a variation of Bug 2, in
which each CPU remains aware that quiescent states are required, but
incorrectly believes that it has already reported a quiescent state
for the current grace period.
To accomplish this,
in \co{__note_gp_changes()}, we clear \co{rnp->qsmask} by adding a statement
\co{rnp->qsmask &= ~rdp->grpmask;} in the last \co{if} code block (line 1739 in \co{tree.c}). 
Then function \co{rcu_report_qs_rnp()} never walks up the \co{rcu_node}
tree, resulting in a liveness violation as in Bug~2.

\paragraph*{Bug 4}
This bug-injection scenario is an alternative code change that gets the
same effect as does Bug~2.
For this alternative,
in \co{__note_gp_changes()}, we set the \co{rdp->qs_pending} field to 0 directly 
(line 1749 in \co{tree.c}). This is a variant of Bug 2 and thus also
a liveness violation.

\paragraph*{Bug 5}
In this bug-injection scenario, CPUs remain aware of the need for
quiescent states.
However, CPUs are prevented from recording their
quiescent states, thus preventing grace periods from ever completing.
To accomplish this, we modify
function \co{rcu_sched_qs()} to return immediately (line 246 in \co{tree.c}),
so that quiescent states are not recorded.
Grace periods therefore never complete, which constitutes a liveness
violation similar to Bug~2.

\paragraph*{Bug 6}
In this bug-injection scenario, CPUs are aware of the need for quiescent
states, and they also record them locally.
However, they are prevented from reporting them up the \co{rcu_node}
tree, which again prevents grace periods from ever completing.
This bug modifies
function \co{rcu_report_qs_rnp()} to return immediately (line 2227 in \co{tree.c}). 
This prevents RCU from walking up the \co{rcu_node} tree, thus preventing grace
periods from ending.
This is again a liveness violation similar to Bug~2.

\paragraph*{Bug 7}
Where Bug~6 prevents quiescent states from being reported up the
\co{rcu_node} tree, this bug-injection scenario causes quiescent
states to be reported up the tree prematurely, before all the
CPUs covered by a given subtree have all reported quiescent states.
To this end,
in \co{rcu_report_qs_rnp()}, we remove the \co{if}-block checking for
\co{rnp->qsmask != 0 || rcu_preempt_blocked_readers_cgp(rnp)} (line 2251 in \co{tree.c}). 
Then the tree-walking process will not stop until it reaches the root, resulting in 
too-short grace periods.
This is therefore a safety violation similar to Bug~1.

\noindent Bugs~2 and~3 would result in a too-short grace period given quiescent-state forcing, 
but such forcing falls outside the scope of this paper.

\subsection{Validating the RCU Model in User-Space}\label{sec:test_model}

\begin{table*}[tb]
\centering
\scalebox{0.85}{%
\begin{tabular}{|l|rr|rr|rr|c|r|c|} \hline
 &
\multicolumn{2}{c|}{\textbf{Successful}} &
\multicolumn{2}{c|}{\textbf{Failing}} &
\multicolumn{2}{c|}{\textbf{}} &
\multicolumn{1}{c|}{\textbf{Max}} &
 &
\\
\multicolumn{1}{|c|}{\textbf{Scenario}} &
\multicolumn{2}{c|}{\textbf{Runs}} &
\multicolumn{2}{c|}{\textbf{Runs}} &
\multicolumn{2}{c|}{\textbf{Timeouts}} &
\multicolumn{1}{c|}{\textbf{VM}} &
\multicolumn{1}{c|}{\textbf{Time}} &
\multicolumn{1}{c|}{\textbf{Result}} \\ \hline
Prove       & 1,000 & (100.0\%) &     0 & (0.0\%)   &     0 & (0.0\%)   & 362\,MB &       4m & Safe                      \\ \hline 
Prove-GP    &     0 & (0.0\%)   & 1,000 & (100.0\%) &     0 & (0.0\%)   & 362\,MB &       5m & GP Completed              \\ \hline 
Bug 1       &   461 & (46.1\%)  &   539 & (53.9\%)  &     0 & (0.0\%)   & 362\,MB &       6m & Assertion Violated        \\ \hline
Bug 2       &     0 & (0.0\%)   &     0 & (0.0\%)   & 1,000 & (100.0\%) & 362\,MB & 16h\,40m & GP Hung                   \\ \hline
Bug 3       &     0 & (0.0\%)   &     0 & (0.0\%)   & 1,000 & (100.0\%) & 362\,MB & 16h\,40m & GP Hung                   \\ \hline
Bug 4       &     0 & (0.0\%)   &     0 & (0.0\%)   & 1,000 & (100.0\%) & 362\,MB & 16h\,40m & GP Hung                   \\ \hline
Bug 5       &     0 & (0.0\%)   &     0 & (0.0\%)   & 1,000 & (100.0\%) & 362\,MB & 16h\,40m & GP Hung                   \\ \hline
Bug 6       &     0 & (0.0\%)   &     0 & (0.0\%)   & 1,000 & (100.0\%) & 362\,MB & 16h\,40m & GP Hung                   \\ \hline
Bug 7 (1R)  &     0 & (0.0\%)   &     0 & (0.0\%)   & 1,000 & (100.0\%) & 362\,MB & 16h\,40m & \mkcol{Safe (Bug Missed)} \\ \hline
Bug 7 (2R)  &   758 & (75.8\%)  &   242 & (24.2\%)  &     0 & (0.0\%)   & 370\,MB &       5m & Assertion Violated        \\ \hline 
\end{tabular}
}
\caption{Experimental Results of Testing the RCU Model in User-Space}
\label{tab:results_run}
\end{table*}

To validate our RCU model before performing verification using CBMC, 
we executed it in user space. We performed 1000 runs for each scenario 
in Section~\ref{sec:bug_cases} using a 60\,s timeout to wait for the end of a 
grace period and a random delay between 0 to 1\,s in the RCU reader task.

The results are reported in Table~\ref{tab:results_run}.  Column 1 
gives the verification scenarios. 
Scenario Prove tests our RCU model without bug injection. Scenario Prove-GP 
tests a weak form of liveness by replacing Figure~\ref{fig:verify_rcu_gp}'s 
assertion with \co{assert(0)} as described in Section~\ref{sec:bug_cases}.
The next three columns present the number and the percentage of successful, 
failing, and timeout runs, respectively. The following two columns give the 
maximum memory consumption and the total runtime. The last column explains 
the results. 

As expected, for scenario Prove, the user program ran to completion
successfully in all runs.  For Prove-GP, it was able to detect the end of a
grace period by triggering an assertion violation in all the runs.  For
Bug~1, an assertion violation was triggered in 559 out of 1000 runs.  
For Bugs 2--6, the user program timed out in all the runs, 
thus a grace period did not complete. 
For Bug 7 with one reader thread, the testing harness failed to
trigger an assertion violation.  However, we were able to observe a failure in
242 out of 1000 runs with two reader threads.



\subsection{Getting CBMC to work on Tree RCU} \label{sec:cbmc_on_rcu}

We have found that getting CBMC to work on our RCU model is non-trivial due
to Tree RCU's complexity combined with CBMC's bit-precise verification.  In
fact, early attempts resulted in SAT formulas that were so large that CBMC
ran out of memory.  After the optimizations described in the remainder of
this section, the largest formula contained around 90 million variables
and 450 million clauses, which enabled CBMC to run to completion.
%

First, instead of placing the scheduling-clock interrupt in its own thread,
we invoke functions \co{rcu_note_context_switch()} and
\co{rcu_process_callbacks()} directly, as described in
Section~\ref{sec:model_irq}.  Also, we invoke
\co{__note_gp_changes()} from \co{rcu_gp_init()} to notify each CPU of a new
grace period, instead of invoking \co{rcu_process_callbacks()}.
%

Second, the support for linked lists in CBMC version 5.4 is limited,
resulting in unreachable code in CBMC's symbolic execution. Thus, we
stubbed all the list-related code in our RCU model, including those for
callback handling.

Third, CBMC's structure-pointer and array encodings result in large formulas
and long formula-generation times.  Our focus on the RCU-sched flavor
allowed us to eliminate RCU-BH's data structures and trivialize the
\co{for_each_rcu_flavor()} flavor-traversal loops.  Our focus on small
numbers of CPUs meant that RCU-sched's \co{rcu_node} tree contained only a
root node, so we also trivialized the \co{rcu_for_each_node_breadth_first()}
loops traversing this tree.

Fourth, CBMC unwinds each loop to the depth specified in its command line
option \co{--unwind}, even when the actual loop depth is smaller.  This
unnecessarily increases formula size, especially for loops containing
intricate RCU code. Since loops in our model can be decided at compile time, 
we therefore used the command line option \co{--unwindset} to specify 
unwinding depths for each individual loop.

Finally, since our test harness only requires one \co{rcu_node} structure
and two \co{rcu_data} structures, we can use 32-bit encodings for \co{int},
\co{long}, and pointers by using the command line option \co{--ILP32}.  This
reduces CBMC's formula size by half compared to the 64-bit default.

\subsection{Results and Discussion}

Table~\ref{tab:results_cbmc} presents the results of our experiments
applying CBMC version~5.4 to verify our RCU model.
Scenario Prove verifies our RCU model without
bug injection over Sequential Consistency (SC).  We also exercise the model
over the weak memory models TSO and PSO in scenarios Prove-TSO and Prove-PSO, 
respectively.  Scenario Prove-GP performs the same reachability check as in 
Section~\ref{sec:test_model} over SC.
We perform the same reachability verification over TSO
and PSO in scenarios Prove-GP-TSO and Prove-GP-PSO, respectively.  Scenarios
Bug~1--7 are the bug-injection scenarios discussed in
Section~\ref{sec:bug_cases}, and are verified over SC, TSO and PSO.  Columns
2--4 give the number of constraints (symbolic program expressions and
partial orders), variables, and clauses of the generated
formula.  The next three columns give the maximum (virtual) memory
consumption, solver runtime, and total runtime of our experiments.  The
final column gives the verification result.


\begin{table*}[tb]
\centering
\scalebox{0.85}{%
\begin{tabular}{|l|c|c|c|c|r|r|c|} \hline
 & & & &
\multicolumn{1}{c|}{\textbf{Max}} &
\multicolumn{1}{c|}{\textbf{Solver}} &
\multicolumn{1}{c|}{\textbf{Total}} &
\\
\multicolumn{1}{|c|}{\textbf{Scenario}} &
\multicolumn{1}{c|}{\textbf{\#Const}} &
\multicolumn{1}{c|}{\textbf{\#Variable}} &
\multicolumn{1}{c|}{\textbf{\#Clause}} &
\multicolumn{1}{c|}{\textbf{VM}} &
\multicolumn{1}{c|}{\textbf{Time}} &
\multicolumn{1}{c|}{\textbf{Time}} &
\multicolumn{1}{c|}{\textbf{Result}} \\ \hline
Prove             &  5.2m & 30.0m & 149.7m & 23\,GB &  9h\,24m &  9h\,36m & Safe \\ \hline
Prove-TSO         &  5.6m & 42.0m & 210.7m & 34\,GB & 10h\,51m & 11h\,4m  & Safe \\ \hline
Prove-PSO         &  5.6m & 41.3m & 207.0m & 34\,GB & 11h\,23m & 11h\,36m & Safe \\ \hline
Prove-GP          &  5.4m & 30.6m & 152.7m & 24\,GB &  3h\,52m &  4h\,5m  & GP Completed \\ \hline
Prove-GP-TSO      &  5.6m & 42.0m & 210.7m & 34\,GB & 13h\,1m  & 13h\,14m & GP Completed \\ \hline
Prove-GP-PSO      &  5.6m & 41.3m & 207.0m & 34\,GB &  8h\,24m &  8h\,37m & GP Completed \\ \hline
Bug 1             &  1.3m & 11.7m &  56.0m &  8\,GB &      31m &      33m & Assertion Violated \\ \hline
Bug 1-TSO         &  1.5m & 17.1m &  83.3m & 13\,GB &      53m &      56m & Assertion Violated \\ \hline
Bug 1-PSO         &  1.5m & 16.5m &  80.4m & 12\,GB &      46m &      48m & Assertion Violated \\ \hline
Bug 2             &  5.2m & 30.0m & 149.6m & 23\,GB &  4h\,25m &  4h\,37m & GP Hung \\ \hline
Bug 2-TSO         &  5.6m & 42.0m & 210.5m & 34\,GB &  9h\,57m & 10h\,10m & GP Hung \\ \hline
Bug 2-PSO         &  5.6m & 41.2m & 206.9m & 34\,GB &  8h\,51m &  9h\,4m  & GP Hung \\ \hline
Bug 3             &  6.3m & 34.8m & 174.1m & 28\,GB &  7h\,11m &  7h\,25m & GP Hung \\ \hline
Bug 3-TSO         &  6.8m & 48.7m & 245.1m & 41\,GB & 19h\,40m & 19h\,55m & GP Hung \\ \hline
Bug 3-PSO         &  6.7m & 48.0m & 241.2m & 41\,GB & 19h\,19m & 19h\,35m & GP Hung \\ \hline
Bug 4             &  4.8m & 27.8m & 138.1m & 22\,GB &  4h\,3m  &  4h\,14m & GP Hung \\ \hline
Bug 4-TSO         &  5.1m & 38.4m & 192.6m & 31\,GB &  8h\,18m &  8h\,30m & GP Hung \\ \hline
Bug 4-PSO         &  5.1m & 37.7m & 188.9m & 31\,GB &  8h\,14m &  8h\,26m & GP Hung \\ \hline
Bug 5             &  5.1m & 29.5m & 146.7m & 23\,GB &  4h\,6m  &  4h\,18m & GP Hung \\ \hline
Bug 5-TSO         &  5.5m & 41.2m & 206.5m & 34\,GB &  5h\,46m &  5h\,59m & GP Hung \\ \hline
Bug 5-PSO         &  5.4m & 40.5m & 202.9m & 33\,GB &  5h\,42m &  5h\,55m & GP Hung \\ \hline
Bug 6             &  1.4m & 13.1m &  63.3m &  9\,GB &      19m &      21m & GP Hung \\ \hline
Bug 6-TSO         &  1.5m & 17.2m &  84.1m & 13\,GB &  1h\,32m &  1h\,33m & GP Hung \\ \hline
Bug 6-PSO         &  1.5m & 16.7m &  81.4m & 12\,GB &  1h\,22m &  1h\,24m & GP Hung \\ \hline
Bug 7 (1R)        &  5.0m & 29.2m & 145.3m & 23\,GB &  8h\,48m &  9h      & \mkcol{Safe (Bug Missed)} \\ \hline
Bug 7-TSO (1R)    &  5.2m & 40.1m & 200.8m & 32\,GB & 11h\,6m  & 11h\,18m & Assertion Violated \\ \hline
Bug 7-PSO (1R)    &  5.1m & 39.4m & 197.2m & 32\,GB & 11h\,32m & 11h\,44m & Assertion Violated \\ \hline
Bug 7 (2R) $^*$
                  & 15.1m & 71.2m & 359.0m & 59\,GB & 19h\,2m  & 19h\,40m & Assertion Violated \\ \hline
Bug 7-TSO (2R) $^*$
                  & 15.6m & 90.4m & 456.9m & 75\,GB & 78h\,12m & 78h\,53m & Assertion Violated \\ \hline
Bug 7-PSO (2R) $^*$
                  & 15.6m & 89.3m & 451.6m & 75\,GB & 84h\,21m & 85h\,2m  & \mkcol{Out of Memory} \\ \hline
\end{tabular}
}
\vspace*{0.03cm} \\
\raggedright
\centering\footnotesize * This experiment was performed on a 64-bit machine running Linux 3.19.8 with twelve Intel Xeon 2.40\,GHz cores and 96\,GB of main memory
\vspace*{0.05cm}
\caption{Experimental Results of CBMC}
\label{tab:results_cbmc}
\end{table*}

Since Tree RCU's implementation in the Linux kernel is sophisticated, its test suite
is non-trivial~\cite{PaulMcKenney2005rcutorture}, comprising several thousand
lines of code.
Therefore, it comes as little surprise that its verification is 
challenging.

In our experiments, CBMC returned all the expected results except
for Bug~7, for which it failed to report a violation of the
assertion \co{assert(r2 == 0 || r1 == 1)} with one RCU reader thread running
over SC.  This failure was due to the approximation of the
scheduling-clock interrupt by a direct function call,
as described in Section~\ref{sec:model_rcu}. 
However, CBMC did report a violation of the assertion
either when two RCU reader threads were present or when run over
TSO or PSO.
All of these cases decrease determinism,
which in turn more faithfully model non-deterministic
scheduling-clock interrupts, allowing the assertion to be violated.

CBMC took more than 9 hours to verify our model over SC (scenario Prove). 
The resulting SAT formulas have more than 5m
constraints, 30m variables and 149m clauses, and occupy
23\,GB of memory.  The formulas for scenarios Prove-TSO and
Prove-PSO are about
40\% larger than the scenario Prove.  They have more than 40m
variables and 200m clauses, and took more than 11 hours and 33\,GB
memory to solve.
Although this verification consumed considerable memory
and CPU, it verified all possible executions and
reorderings permitted by TSO and PSO,
a tiny subset of which are reached by
the \co{rcutorture} test suite.

CBMC proved that grace periods can end (i.e., \co{assert(0)} is
violated), over SC (Prove-GP), TSO (Prove-GP-TSO), and PSO
(Prove-GP-PSO).  The sizes of resulting formulas and memory consumption
are similar to those of the three Prove scenarios.
However, it took CBMC only about 4, 13, and 8.5 hours to find an
violation of \co{assert(0)} in Prove-GP, Prove-GP-TSO, and Prove-GP-PSO, respectively.

For the bug-injection scenarios described in Section~\ref{sec:bug_cases}, CBMC
was able to return the expected results in all scenarios over SC except for Bug~7,
as noted earlier.
The formula size varies from scenario to
scenario, with 27m--35m variables and 138m--174m
clauses.  The runtime was 4--9 hours and memory consumption exceeded
22\,GB.  The exceptions are Bugs~1 and 6, which have fewer than
14m variables and 64m clauses, and took less than 35 mins and
about 9\,GB of memory to solve.  This reduction was due to the large amount
of code removed by the bug injections in these scenarios.

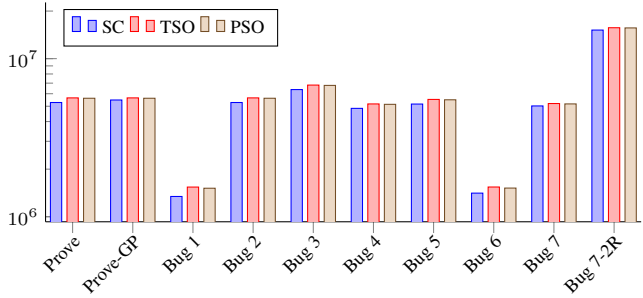
\begin{figure}[tbp]
\centering
\captionsetup{justification=centering}
\begin{tikzpicture}
\scriptsize
\begin{axis}[
  ybar,
  bar width=0.15cm,
  height=4.5cm,
  width=9.5cm,
  axis lines*=left, 
  ymode=log,
  symbolic x coords={Prove, Prove-GP, Bug 1, Bug 2, Bug 3, 
                     Bug 4, Bug 5, Bug 6, Bug 7, Bug 7-2R,
                    },
  xtick=data,
  xticklabel style={
    inner sep=0pt,
    anchor=north east,
    rotate=45
  },
  enlarge y limits=0.15, 
  enlarge x limits=0.05, 
  legend style={
    legend pos=north west,
    legend columns=-1,
    font=\scriptsize},
]

\addplot 
  coordinates {(Prove, 5279600) (Prove-GP, 5476540)
               (Bug 1, 1343449) (Bug 2, 5279584) (Bug 3, 6374373)
               (Bug 4, 4847980) (Bug 5, 5161874) (Bug 6, 1410495)
               (Bug 7, 5022249) (Bug 7-2R, 15165557)
              };

\addplot 
  coordinates {(Prove, 5646959) (Prove-GP, 5646940)
               (Bug 1, 1540645) (Bug 2, 5646940) (Bug 3, 6805631)
               (Bug 4, 5170928) (Bug 5, 5522168) (Bug 6, 1541937)
               (Bug 7, 5201744) (Bug 7-2R, 15691102)
              };

\addplot 
  coordinates {(Prove, 5617154) (Prove-GP, 5617135)
               (Bug 1, 1514657) (Bug 2, 5617135) (Bug 3, 6773763)
               (Bug 4, 5141123) (Bug 5, 5492607) (Bug 6, 1518307)
               (Bug 7, 5172720) (Bug 7-2R, 15647504)
              };

\legend{SC, TSO, PSO}
\end{axis}
\end{tikzpicture}
\caption{Number of Constraints in the SAT Formulas}
\label{fig:barchart_sat_constr}
\end{figure}

\begin{figure}[tbp]
\centering
\captionsetup{justification=centering}
\begin{tikzpicture}
\scriptsize
\begin{axis}[
  ybar,
  bar width=0.15cm,
  height=4.5cm,
  width=9.5cm,
  axis lines*=left, 
  ymode=log,
  symbolic x coords={Prove, Prove-GP, Bug 1, Bug 2, Bug 3, 
                     Bug 4, Bug 5, Bug 6, Bug 7, Bug 7-2R,
                    },
  xtick=data,
  xticklabel style={
    inner sep=0pt,
    anchor=north east,
    rotate=45
  },
  enlarge y limits=0.15, 
  enlarge x limits=0.05, 
  legend style={
    legend pos=north west,
    legend columns=-1,
    font=\scriptsize},
]

\addplot 
  coordinates {(Prove, 30085337) (Prove-GP, 30655428)
               (Bug 1, 11719966) (Bug 2, 30056615) (Bug 3, 34856577)
               (Bug 4, 27804363) (Bug 5, 29510828) (Bug 6, 13165176)
               (Bug 7, 29242760) (Bug 7-2R, 71205400)
              };

\addplot 
  coordinates {(Prove, 42042386) (Prove-GP, 42041740)
               (Bug 1, 17120555) (Bug 2, 42013372) (Bug 3, 48788433)
               (Bug 4, 38480891) (Bug 5, 41239083) (Bug 6, 17286058)
               (Bug 7, 40139251) (Bug 7-2R, 90444903)
              };

\addplot 
  coordinates {(Prove, 41327066) (Prove-GP, 41326420)
               (Bug 1, 16548819) (Bug 2, 41298052) (Bug 3, 48023601)
               (Bug 4, 37765571) (Bug 5, 40529619) (Bug 6, 16766198)
               (Bug 7, 39442675) (Bug 7-2R, 89398551)
              };

\legend{SC, TSO, PSO}
\end{axis}
\end{tikzpicture}
\caption{Number of Variables in the SAT Formulas}
\label{fig:barchart_sat_var}
\end{figure}
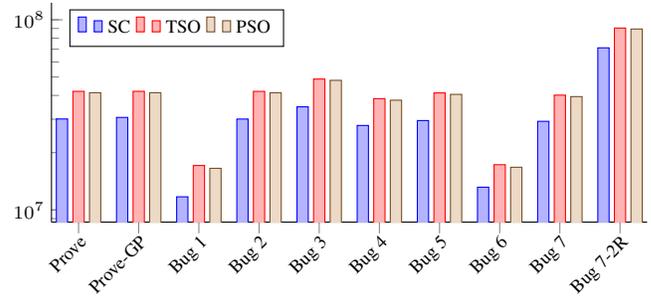

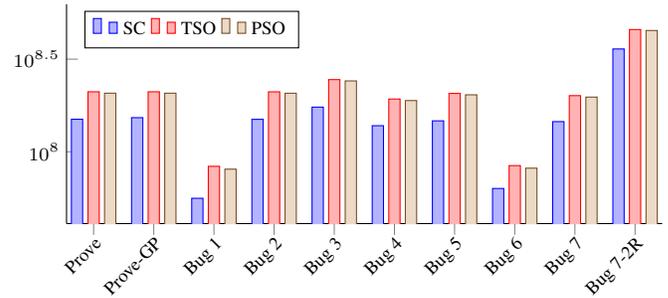
\begin{figure}[tbp]
\centering
\captionsetup{justification=centering}
\begin{tikzpicture}
\scriptsize
\begin{axis}[
  ybar,
  bar width=0.15cm,
  height=4.5cm,
  width=9.5cm,
  axis lines*=left, 
  ymode=log,
  symbolic x coords={Prove, Prove-GP, Bug 1, Bug 2, Bug 3, 
                     Bug 4, Bug 5, Bug 6, Bug 7, Bug 7-2R,
                    },
  xtick=data,
  xticklabel style={
    inner sep=0pt,
    anchor=north east,
    rotate=45
  },
  enlarge y limits=0.15, 
  enlarge x limits=0.05, 
  legend style={
    legend pos=north west,
    legend columns=-1,
    font=\scriptsize},
]

\addplot 
  coordinates {(Prove, 149758548) (Prove-GP, 152743545)
               (Bug 1, 56027980) (Bug 2, 149643492) (Bug 3, 174131331)
               (Bug 4, 138197043) (Bug 5, 146787005) (Bug 6, 63302559)
               (Bug 7, 145389516) (Bug 7-2R, 359021922)
              };

\addplot 
  coordinates {(Prove, 210708442) (Prove-GP, 210705615)
               (Bug 1, 83392397) (Bug 2, 210592015) (Bug 3, 245157184)
               (Bug 4, 192605939) (Bug 5, 206569643) (Bug 6, 84131818)
               (Bug 7, 200857404) (Bug 7-2R, 456973933)
              };

\addplot 
  coordinates {(Prove, 207042629) (Prove-GP, 207039802)
               (Bug 1, 80481851) (Bug 2, 206926202) (Bug 3, 241237629)
               (Bug 4, 188940126) (Bug 5, 202933839) (Bug 6, 81485361)
               (Bug 7, 197287644) (Bug 7-2R, 451611664)
              };

\legend{SC, TSO, PSO}
\end{axis}
\end{tikzpicture}
\caption{Number of Clauses in the SAT Formulas}
\label{fig:barchart_sat_clause}
\end{figure}

\begin{figure}[tbp]
\centering
\captionsetup{justification=centering}
\begin{tikzpicture}
\scriptsize
\begin{axis}[
  ybar,
  bar width=0.15cm,
  height=4.5cm,
  width=9.5cm,
  axis lines*=left, 
  ymode=log,
  symbolic x coords={Prove, Prove-GP, Bug 1, Bug 2, Bug 3, 
                     Bug 4, Bug 5, Bug 6, Bug 7, Bug 7-2R,
                    },
  xtick=data,
  xticklabel style={
    inner sep=0pt,
    anchor=north east,
    rotate=45
  },
  enlarge y limits=0.15, 
  enlarge x limits=0.05, 
  legend style={
    legend pos=north west,
    legend columns=-1,
    font=\scriptsize},
]

\addplot 
  coordinates {(Prove, 34570.5) (Prove-GP, 14698.4)
               (Bug 1, 2002.53) (Bug 2, 16644.8) (Bug 3, 26716.8)
               (Bug 4, 15231.3) (Bug 5, 15490.8) (Bug 6, 1255.06)
               (Bug 7, 32373.1) (Bug 7-2R, 70822.7)
              };

\addplot 
  coordinates {(Prove, 39820.1) (Prove-GP, 47643.6)
               (Bug 1, 3360.95) (Bug 2, 36608.3) (Bug 3, 71708.5)
               (Bug 4, 30599.2) (Bug 5, 21550.4) (Bug 6, 5608.83)
               (Bug 7, 40667.7) (Bug 7-2R, 283950)
              };

\addplot 
  coordinates {(Prove, 41754.8) (Prove-GP, 31002.2)
               (Bug 1, 2902.64) (Bug 2, 32616.8) (Bug 3, 70477.1)
               (Bug 4, 30355) (Bug 5, 21296.9) (Bug 6, 5040.28)
               (Bug 7, 42226) (Bug 7-2R, 306133)
              };

\legend{SC, TSO, PSO}
\end{axis}
\end{tikzpicture}
\caption{Total Runtime in Seconds}
\label{fig:barchart_runtime}
\end{figure}
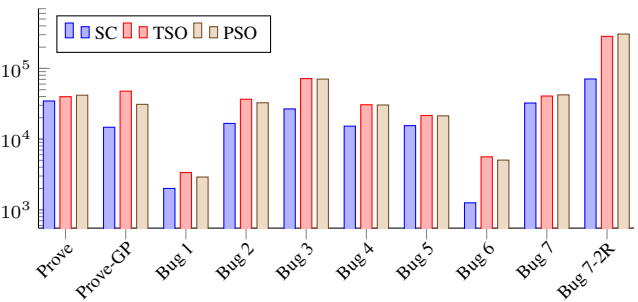

\begin{figure}[tbp]
\centering
\captionsetup{justification=centering}
\begin{tikzpicture}
\scriptsize
\begin{axis}[
  ybar,
  bar width=0.15cm,
  height=4.5cm,
  width=9.5cm,
  axis lines*=left, 
  ymode=log,
  symbolic x coords={Prove, Prove-GP, Bug 1, Bug 2, Bug 3, 
                     Bug 4, Bug 5, Bug 6, Bug 7, Bug 7-2R,
                    },
  xtick=data,
  xticklabel style={
    inner sep=0pt,
    anchor=north east,
    rotate=45
  },
  enlarge y limits=0.15, 
  enlarge x limits=0.05, 
  legend style={
    legend pos=north west,
    legend columns=-1,
    font=\scriptsize},
]

\addplot 
  coordinates {(Prove, 23.27) (Prove-GP, 23.90)
               (Bug 1, 8.24) (Bug 2, 23.26) (Bug 3, 28.04)
               (Bug 4, 22.18) (Bug 5, 23.02) (Bug 6, 9.03)
               (Bug 7, 22.87) (Bug 7-2R, 59.07)
              };

\addplot 
  coordinates {(Prove, 34.00) (Prove-GP, 34.00)
               (Bug 1, 12.60) (Bug 2, 34.01) (Bug 3, 41.18)
               (Bug 4, 31.49) (Bug 5, 33.65) (Bug 6, 12.59)
               (Bug 7, 31.93) (Bug 7-2R, 74.80)
              };

\addplot 
  coordinates {(Prove, 33.76) (Prove-GP, 33.76)
               (Bug 1, 12.42) (Bug 2, 33.75) (Bug 3, 40.95)
               (Bug 4, 31.27) (Bug 5, 33.04) (Bug 6, 12.44)
               (Bug 7, 31.71) (Bug 7-2R, 74.51)
              };

\legend{SC, TSO, PSO}
\end{axis}
\end{tikzpicture}
\caption{Maximum Memory Consumption in Gigabytes}
\label{fig:barchart_memory}
\end{figure}
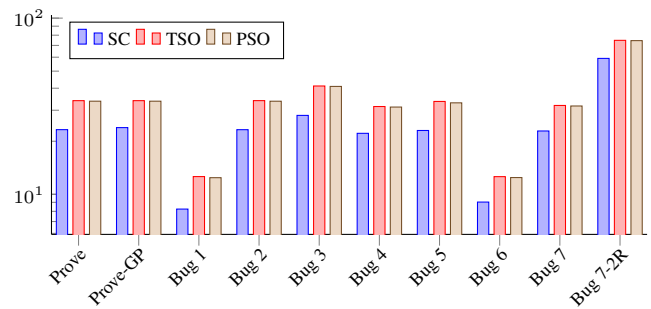

Figures~\ref{fig:barchart_sat_constr}--\ref{fig:barchart_sat_clause} 
compare the formula size between SC, TSO and TSO. Comparison of 
runtime and memory can be found in Figures~\ref{fig:barchart_runtime} 
and \ref{fig:barchart_memory}. As we can see,
the runtime and memory overhead for the TSO and PSO variants of a given experiment 
are quite similar.
The overheads of TSO
are slightly higher than those of PSO in all bug-injection scenarios except
for Bug 7 on which PSO had longer runtime.
However, the overhead of TSO and PSO is significantly larger than that of SC,
with up to 340\% (Bug 6 runtime) and 50\% (Bug 1 memory) increases.
The runtime was 5--19 hours and memory consumption exceeded
31\,GB in all scenarios except Bug 1 and 6.  The numbers of variables and
clauses are 37m--49m and 188m--245m, respectively, around 130\%
greater than SC.

The two-reader variant of Bug~7 has by far the longest
runtime, consuming more than 19~hours and
78~hours over SC and TSO, respectively, comparing to 9~hours and 11~hours
with one reader.  It also consumed about 75\,GB memory, more than double
the one-reader variant.  For PSO, with two reader threads CBMC's solver ran
out of memory after 85~hours whereas with one reader it completed
in less than 12~hours.  The increased overhead
is due to the
additional RCU reader's call to
\co{rcu_process_callbacks()}.
This in turn results in
more than a 125\% increase in the number of constraints, variables, and
clauses.  For example, the two-reader TSO formula has
triple the constraints and double the variables and clauses
of the one-reader case.
\section{Related Work}
%
McKenney has applied the SPIN model checker to verify RCU's \co{NO_HZ_FULL_SYSIDLE}
functionality \cite{VerificationChallenges},
and interactions between dyntick-idle and 
non-maskable interrupts~\cite{ValDyntickNMI}. 
Desnoyers et al.~\cite{DesnoyersOSR13} propose a virtual architecture 
to model out-of-order memory accesses and instruction scheduling.
User-level RCU \cite{DesnoyersTPDS12UserRCU} is modeled and verified 
in the proposed architecture using the SPIN model checker. 

These efforts require an error-prone translation from C to SPIN's 
modeling language, and therefore are not appropriate for regression testing. 
By contrast, our work constructs an RCU model directly from its 
source code from the Linux kernel, and verifies it using automated 
verification tool. 

Alglave et al.~\cite{AlglaveCAV13} introduce a symbolic encoding 
for verifying concurrent software over a range of memory models 
including SC, TSO and PSO.
They implement the encoding in the CBMC bounded model checker and 
use the tool to verify \co{rcu_assign_pointer()} and \co{rcu_dereference()}.

McKenney has used CBMC to verify Tiny
RCU~\cite{VerificationChallenges}, a trivial Linux-kernel
RCU implementation for uni-core systems.
Roy has applied the same tool to verify a significant portion of Sleepable RCU (SRCU).
CBMC is now part of the regression test suite of SRCU in the Linux kernel~\cite{RoyValSRCU17}.

Concurrently with our work, Kokologiannakis et al.~verify Tree RCU
using Nidhugg~\cite{NidhuggTreeRCUSPIN17}. Since Nidhugg has better
list support and does not model data non-determinism, they are able
to verify more scenarios with less CPU and memory consumption.
But some portions of RCU use atomic read-modify-write operations that
can give nondeterministic results. So we hypothesize that
data non-determinism will be required to verify RCU's dyntick-idle and
\co{rcu_barrier()} components.

Groce et al.~\cite{GroceASE15RCU} introduce a falsification-driven
verification methodology that is based on a variation of mutation 
testing. By using CBMC, they are able to find two holes in 
\co{rcutorture}--RCU's stress testing suite, one of which was hiding 
a real bug in Tiny RCU.
Further work on real hardware has identified two more
\co{rcutorture} holes, one of which was hiding a real bug in Tasks
RCU~\cite{JonathanCorbet2014RCU-tasks} and the other of which was hiding
a minor performance bug in Tree RCU.

In this work, we use CBMC to verify the implementation of Linux-kernel
Tree RCU for multi-core systems, which is more complex
and sophisticated, over SC, TSO, and PSO.

Gotsman et al.~\cite{YangESOP13RCU} use an extended concurrent separation logic 
to formalize the concept 
of grace period and prove an abstract implementation of RCU over SC.
Tassarotti et al.~\cite{DreyerPLDI15RCU} use GPS, a recently 
developed program logic for the C/C++11 memory model, to carry out 
a formal proof of a simple implementation of user-level 
RCU for a singly-linked list. They assuma the ``release-acquire'' semantics, 
which is weaker than SC but stronger than memory models used by
real-world RCU implementations.
These formal proofs are performed manually on simple implementations 
of RCU. By contrast, our work applies an automated verification tool with a 
test harness to verify the grace-period property of a real-world
implementation of RCU over SC, TSO, and PSO.

Formal verification has started to make its way into real-world practice 
of verifying large non-trivial code bases. Cal\-ca\-gno et al.~\cite{CalcagnoNASA15} 
describe integrating a static-analysis tool into Facebook's software development cycle.
Malacaria et al.~\cite{MalacariaISOLA16} apply CBMC to analyze information leakage
of the OpenSSL implementation. We believe that our work is an important step towards 
integration of verification into Linux-kernel RCU's regression test suite.
\section{Conclusion}
This paper overviews the implementation of Tree RCU in the 
Linux Kernel, and describes how to construct a model directly from
its source code. It then shows how to use the CBMC model checker to 
verify a significant part of the Tree RCU implementation automatically, 
which to the best of our knowledge is unprecedented.
This work demonstrates that RCU is a rich example to drive research:
it is small enough to provide models that can just
barely be verified by existing tools, but it also has sufficient concurrency
and complexity to drive significant advances in techniques and tooling.

For future work, we plan to 
add quiescent-state forcing and grace-period expediting into our model 
and verify their safety and liveness properties, using more sophisticated 
test harnesses that pass through multiple grace periods and operate on 
a larger tree structure.
We also plan to model and verify the preemptible version of Tree RCU, 
which we expect to be quite challenging. Moreover, there is much fertile ground 
verifying uses of RCU in the Linux kernel, for example, the Virtual File
System (VFS).
%

There are also potential improvements for CBMC to better support future
RCU verification efforts. For instance, better support of lists is required
to verify RCU's callback handling mechanism. A field-sensitive SSA encoding
for structures and a thread-aware slicer will help reduce encoding size,
and therefore improve scalability.

This work demonstrates the nascent ability of SAT-based
formal-verification tools to handle real-world pro\-duc\-tion-quality
synchronization primitives, as exemplified by Linux-kernel
Tree RCU on weakly ordered TSO and PSO systems.
Although modeling weak ordering incurs a significant performance
penalty, this penalty is not excessive.
We therefore hypothesize that use of these tools for highly concurrent
multithreaded software
will reach mainstream within 3-5 years, especially given recent
rates of improvement.
\bibliographystyle{abbrvnat} 
\raggedright
\bibliography{paper}

\begin{thebibliography}{24}
\providecommand{\natexlab}[1]{#1}
\providecommand{\url}[1]{\texttt{#1}}
\expandafter\ifx\csname urlstyle\endcsname\relax
  \providecommand{\doi}[1]{doi: #1}\else
  \providecommand{\doi}{doi: \begingroup \urlstyle{rm}\Url}\fi

\bibitem[Lin()]{LinuxKernel}
The {Linux} kernel.
\newblock \url{https://www.kernel.org/}.

\bibitem[Lin(2013)]{LinuxServerNum13}
Where is the {Internet}?
\newblock
  \url{http://www.whoishostingthis.com/blog/2013/12/06/internet-infographic/},
  December 2013.

\bibitem[Alglave et~al.(2013)Alglave, Kroening, and Tautschnig]{AlglaveCAV13}
J.~Alglave, D.~Kroening, and M.~Tautschnig.
\newblock Partial orders for efficient bounded model checking of concurrent
  software.
\newblock In \emph{Proceedings of the 25th International Conference on Computer
  Aided Verification (CAV)}, volume 8044 of \emph{LNCS}, pages 141--157.
  Springer, 2013.

\bibitem[Burch et~al.(1992)Burch, Clarke, McMillan, Dill, and
  Hwang]{BurchInfComput92}
J.~R. Burch, E.~M. Clarke, K.~L. McMillan, D.~L. Dill, and L.~J. Hwang.
\newblock Symbolic model checking: $10^{20}$ states and beyond.
\newblock \emph{Information and Computation}, 98\penalty0 (2):\penalty0
  142--170, 1992.

\bibitem[Calcagno et~al.(2015)Calcagno, Distefano, Dubreil, Gabi, Hooimeijer,
  Luca, O'Hearn, Papakonstantinou, Purbrick, and Rodriguez]{CalcagnoNASA15}
C.~Calcagno, D.~Distefano, J.~Dubreil, D.~Gabi, P.~Hooimeijer, M.~Luca, P.~W.
  O'Hearn, I.~Papakonstantinou, J.~Purbrick, and D.~Rodriguez.
\newblock Moving fast with software verification.
\newblock In \emph{Proceedings of the 7th {NASA} Formal Methods Symposium
  (NFM)}, volume 9058 of \emph{LNCS}, pages 3--11. Springer, 2015.

\bibitem[Callaham(2015)]{AndroidNum15}
J.~Callaham.
\newblock Google says there are now 1.4 billion active {Android} devices
  worldwide.
\newblock
  \url{http://www.androidcentral.com/google-says-there-are-now-14-billion-active-}\newline\url{android-devices-worldwide},
  September 2015.

\bibitem[Clarke et~al.(2004)Clarke, Kroening, and Lerda]{KroeningTACAS04CBMC}
E.~M. Clarke, D.~Kroening, and F.~Lerda.
\newblock A tool for checking {ANSI-C} programs.
\newblock In \emph{Proceedings of the 10th International Conference on Tools
  and Algorithms for the Construction and Analysis of Systems (TACAS)}, volume
  2988 of \emph{LNCS}, pages 168--176. Springer, 2004.

\bibitem[Corbet(2014)]{JonathanCorbet2014RCU-tasks}
J.~Corbet.
\newblock The {RCU}-tasks subsystem.
\newblock \url{http://lwn.net/Articles/607117/}, July 2014.

\bibitem[Desnoyers et~al.(2012)Desnoyers, McKenney, Stern, Dagenais, and
  Walpole]{DesnoyersTPDS12UserRCU}
M.~Desnoyers, P.~E. McKenney, A.~S. Stern, M.~R. Dagenais, and J.~Walpole.
\newblock User-level implementations of read-copy update.
\newblock \emph{{IEEE} Transactions on Parallel and Distributed Systems},
  23\penalty0 (2):\penalty0 375--382, 2012.

\bibitem[Desnoyers et~al.(2013)Desnoyers, McKenney, and
  Dagenais]{DesnoyersOSR13}
M.~Desnoyers, P.~E. McKenney, and M.~R. Dagenais.
\newblock Multi-core systems modeling for formal verification of parallel
  algorithms.
\newblock \emph{{ACM} SIGOPS Operating Systems Review}, 47\penalty0
  (2):\penalty0 51--65, 2013.

\bibitem[Gotsman et~al.(2013)Gotsman, Rinetzky, and Yang]{YangESOP13RCU}
A.~Gotsman, N.~Rinetzky, and H.~Yang.
\newblock Verifying concurrent memory reclamation algorithms with grace.
\newblock In \emph{Proceedings of the 22nd European Symposium on Programming
  (ESOP)}, volume 7792 of \emph{LNCS}, pages 249--269. Springer, 2013.

\bibitem[Groce et~al.(2015)Groce, Ahmed, Jensen, and McKenney]{GroceASE15RCU}
A.~Groce, I.~Ahmed, C.~Jensen, and P.~E. McKenney.
\newblock How verified is my code? {Falsification}-driven verification.
\newblock In \emph{Proceedings of the 30th {IEEE/ACM} International Conference
  on Automated Software Engineering (ASE)}, pages 737--748. {IEEE} Computer
  Society, 2015.

\bibitem[Holzmann(1997)]{HolzmannTSE97SPIN}
G.~J. Holzmann.
\newblock The model checker {SPIN}.
\newblock \emph{{IEEE} Transaction of Software Engineering}, 23\penalty0
  (5):\penalty0 279--295, 1997.

\bibitem[Jiangshan(2008)]{LaiJiangshan2008NewClassicAlgorithm}
L.~Jiangshan.
\newblock [{RFC}][{PATCH}] rcu classic: new algorithm for callbacks-processing.
\newblock Available: Linux-kernel git SHA-1 5127bed588a2 [Viewed July 9, 2016],
  June 2008.

\bibitem[Kokologiannakis and Sagonas(2017)]{NidhuggTreeRCUSPIN17}
M.~Kokologiannakis and K.~Sagonas.
\newblock Stateless model checking of the {Linux} kernel’s hierarchical
  {Read-Copy-Update} ({Tree RCU}).
\newblock In \emph{SPIN}, 2017.

\bibitem[Malacaria et~al.(2016)Malacaria, Tautschnig, and
  Distefano]{MalacariaISOLA16}
P.~Malacaria, M.~Tautschnig, and D.~Distefano.
\newblock Information leakage analysis of complex {C} code and its application
  to openssl.
\newblock In \emph{Proceedings of the 7th International Symposium on Leveraging
  Applications of Formal Methods, Verification and Validation (ISOLA)}, volume
  9952 of \emph{LNCS}, pages 909--925, 2016.

\bibitem[McKenney(2005)]{PaulMcKenney2005rcutorture}
P.~E. McKenney.
\newblock {[PATCH]} {RCU} torture testing.
\newblock Available:
  \url{http://lkml.kernel.org/g/20051001182056.GA1613@us.ibm.com} [Viewed July
  9, 2016], October 2005.

\bibitem[McKenney(2008)]{ValDyntickNMI}
P.~E. McKenney.
\newblock Integrating and validating dynticks and preemptable {RCU}.
\newblock \url{https://lwn.net/Articles/279077/}, April 2008.

\bibitem[McKenney(2015)]{VerificationChallenges}
P.~E. McKenney.
\newblock Verification challenges.
\newblock \url{http://paulmck.livejournal.com/tag/verification%20challenge},
  April 2015.

\bibitem[McKenney and Slingwine(1998)]{McKenneyRCU98}
P.~E. McKenney and J.~D. Slingwine.
\newblock Read-copy update: Using execution history to solve concurrency
  problems.
\newblock In \emph{Proceedings of the 10th International Conference on Parallel
  and Distributed Computing and Systems}, pages 509--518, 1998.

\bibitem[McKenney and Walpole(2008)]{McKenneyOSR08}
P.~E. McKenney and J.~Walpole.
\newblock Introducing technology into the {Linux} kernel: a case study.
\newblock \emph{{ACM} SIGOPS Operating Systems Review}, 42\penalty0
  (5):\penalty0 4--17, 2008.

\bibitem[McKenney et~al.(2013)McKenney, Boyd-Wickizer, and
  Walpole]{McKenneyRCUsageReport13}
P.~E. McKenney, S.~Boyd-Wickizer, and J.~Walpole.
\newblock {RCU} usage in the {Linux} kernel: One decade later.
\newblock Technical Report, 2013.

\bibitem[Roy(2017)]{RoyValSRCU17}
L.~Roy.
\newblock rcutorture: Add {CBMC}-based formal verification for {SRCU}.
\newblock \url{https://www.spinics.net/lists/kernel/msg2421833.html}, 2017.

\bibitem[Tassarotti et~al.(2015)Tassarotti, Dreyer, and
  Vafeiadis]{DreyerPLDI15RCU}
J.~Tassarotti, D.~Dreyer, and V.~Vafeiadis.
\newblock Verifying read-copy-update in a logic for weak memory.
\newblock In \emph{Proceedings of the 36th {ACM} {SIGPLAN} Conference on
  Programming Language Design and Implementation (PLDI)}, pages 110--120.
  {ACM}, 2015.

\end{thebibliography}


\end{document}